\documentclass{article}
\usepackage[round]{natbib}
\usepackage{changes}
\usepackage{wrapfig}
\usepackage{url}
\usepackage{bm}
\usepackage{pdflscape}
\usepackage{mwe}
\usepackage{geometry}
\usepackage{verbatim}
\usepackage{float}
\usepackage{siunitx}
\usepackage{doi}
\usepackage{multirow}
\usepackage{longtable}
\usepackage{booktabs}
\usepackage{bm}
\usepackage{svg}
\usepackage{amsfonts}
\usepackage{amsmath}
\usepackage{hyperref}
\usepackage{xcolor}
\usepackage{tikz}
\usepackage{times}
\usepackage{caption}
\usepackage{subcaption}
\usepackage{authblk}
\usepackage[T1]{fontenc}
\usepackage[utf8]{inputenc}
\usepackage{dsfont}
\usepackage{lineno}
\definecolor{turquoise}{cmyk}{0.99,0.70,0.,0.01}
\definecolor{darkgreen}{cmyk}{0.5,0.,0.5,0.5}

\hypersetup{
    colorlinks=true,
    linkcolor=turquoise,
    allcolors=turquoise
}

\DeclareMathOperator{\diag}{diag}

\title{Small bodies global gravity inversion via the level-set method}

\author[1,2]{A. Caldiero \footnote{email: alfonso.caldiero@oma.be}}
\author[1,2]{S. Le Maistre}

\affil[1]{\small Earth and Life Institute, Université Catholique de Louvain, Place Louis Pasteur 3, 1348 Louvain-La-Neuve, Belgium}
\affil[2]{\small Reference Systems and Planetology, Royal Observatory of Belgium, 3 Avenue Circulaire, 1180 Uccle, Belgium}
\date{} 
\begin{document}

\maketitle 
%% Use Roman numerals for the page numbers of the title pages and table of
%% contents.
%% Use Arabic numerals for the page numbers of the chapters.
%% \mainmatter
\tableofcontents
\begin{abstract}
We propose an approach to infer large-scale heterogeneities within a small celestial body from measurements of its gravitational potential, provided for instance by spacecraft radio-tracking. The non-uniqueness of the gravity inversion is here mitigated by limiting the solutions to piecewise-constant density distributions, thus composed of multiple regions of uniform density (mass anomalies) dispersed in a background medium. The boundary of each anomaly is defined implicitly as the 0-level surface of a scalar field (called the level-set function), so that by modifying this field the shape and location of the anomaly are varied. The gravitational potential associated with a density distribution is here computed via a line-integral polyhedron method, yielding the coefficients of its spherical harmonics expansion. The density distribution is then adjusted via an iterative least-squares approach with Tikhonov regularization, estimating at every iteration corrections to the level-set function, the density contrast of each anomaly, and the background density, in order to minimize the residuals between the predicted gravity coefficients and those measured.
Given the non-convexity of the problem and the lack of prior knowledge assumed (save for the shape of the body), the estimation process is repeated for several random initial distributions, and the resulting solutions are clustered based on global properties independent of the input measurements. This provides families of candidate interior models in agreement with the data, and the spread of the local density values across each family is used to assess the uncertainties associated with the estimated distributions. We present multiple synthetic tests with increasingly more realistic settings (in terms of gravity resolution and precision, and of shape, size and distribution of the internal heterogeneities), showing that our method is generally able to retrieve a ground-truth mass distribution even with noisy data. For further validation, we present an application of the method to real data, namely the Bennu gravity coefficients measured by the OSIRIS-REx team. 
\end{abstract}

\section{Introduction}
\label{sec:introduction}
Any information on the current interior structure of a small body can provide insights into its formation, as well as its collisional and dynamical evolution, all strictly related to the history of the solar system itself \citep{walshRubblePileAsteroids2018}. 
Although ground-based methods can allow to constrain the internal mass distribution of asteroids, either in terms of their macroporosities \citep{carryDensityAsteroids2012}, or from their spin evolution \citep{lowryInternalStructureAsteroid2014}, and close-encounters with our planet could prove useful to test candidate mass distributions against their tidal response \citep{dinsmoreConstrainingInteriorsAsteroids2023}, \emph{in-situ} observations remain the one reliable way to determine properties of an asteroid beyond the bulk. Surface geological structures, seismic analysis, tidal dissipation, and radar sounding \citep{kofmanInteriorComet67P2020} can all help infer the distribution of mass within a small body \citep{scheeresAsteroidInteriorsMorphology2015a}, as can the measurement of its gravitational potential. The external gravitational potential is a direct expression of its internal mass distribution, and can be characterized from its perturbation on the motion of spacecraft in its vicinity. The precise reconstruction of a spacecraft orbit from radio-tracking to retrieve such signals (radio science) requires no additional payload other than its telecommunication system, making gravity ubiquitous among the outputs of a space mission to any celestial body. We aim therefore to propose an approach to retrieve candidate mass distributions inside small bodies from their gravity field, as measured in a radio science campaign. \\

Robotic exploration of the Solar System has allowed for several encounters with small bodies, be it in opportunistic flybys or in the frame of dedicated missions. The NEAR Shoemaker mission to asteroid Eros \citep{miller_determination_2002-1} opened an era of small bodies exploration which is far from reaching its peak, with highlights such as the JAXA sample-return missions Hayabusa \citep{fujiwaraRubblePileAsteroidItokawa2006} to asteroid Itokawa and Hayabusa-2 \citep{watanabeHayabusa2ArrivesCarbonaceous2019a} to asteroid Ryugu, the NASA missions Dawn to Vesta and Ceres \citep{konopliv_ceres_2018} and OSIRIS-REx to Bennu \citep{scheeres_heterogeneous_2020}, and the ESA Rosetta spacecraft rendez-vous with comet 67P/Churymov-Gerasimenko \citep{patzoldHomogeneousNucleusComet2016}. 
While, due to the fast decay of the gravity signal with distance, most of these encounters have only allowed to determine the central gravity of the target, and thus its total mass, a handful of missions have successfully measured the extended gravitational potential of their target (see Figure \ref{fig:noiseProfilePlot}), and the growing number of dedicated small bodies missions planned for the coming years could increase this count.
As the OSIRIS-REx spacecraft, fresh from returning to Earth samples from asteroid Bennu, prepares its approach to Apophis, various new encounters are expected within the next decade, with NASA probes Lucy \citep{levisonLucyMissionTrojan2021} and Psyche \citep{zuberPsycheGravityInvestigation2022} already on the way to their targets (the Jupyter Trojans and the metallic asteroid Psyche, respectively).
%, and planned missions like Tianwen-2 and DESTINY+. 
The ESA Hera mission \citep{michelESAHeraMission2022} is expected to reach the Didymos binary system in 2027, already visited by the NASA DART spacecraft in its planetary-defense demonstration of the deflection of the secondary body, Dimorphos. Juventas, one of the two CubeSats on-board the Hera spacecraft (the other being Milani), is planned to land on Dimorphos and carries multiple scientific instruments, including a gravimeter \citep{ritterMeasuringGravityGRASS2022a} and a ground-penetrating radar \citep{heriqueJuRaJuventasRadar2022}, that would provide local measurements complementing the global gravity field in the constraining of the interior structure of the asteroid. Finally, the approach we present here is equally suited to irregular bodies like the Martian moons Phobos and Deimos, which will be targeted by the JAXA MMX mission \citep{matsumotoMMXGeodesyInvestigations2021}.   \\

As is well-known, returning from the measured gravity potential to the internal mass distribution which generates it is an ill-posed problem: attempts to model the interior of a body from gravity measurements have to deal with the non-uniqueness of the solution and its instability \citep{blakelyPotentialTheoryGravity1995, michel_unified_2008, chaoInversionMassDistribution2005}. Unambiguous reconstruction of the interior therefore relies heavily on the injection of \emph{a priori} knowledge. For the Earth, where constraints from other observations (such as seismic or magnetic) are widely available, this has been performed extensively both at global and local scales \citep{zheglova_multiple_2017}, as reflected by the wide range of commercial and open-source tools available \citep[e.g.][]{Ruecker2017}. The interior of the planet is generally approximated by prismatic elements for which the density is estimated via least-squares inversion with various degrees of regularization \citep{li_3-d_1998}, although spectral or Monte-Carlo methods are equally popular \citep{michel_unified_2008}. \\

For small bodies, however, the initial knowledge about the internal structure may be little to none. Therefore, except in special cases where additional observables and theoretical considerations may limit the possible models to a small set of families \citep{le_maistre_signature_2019}, exploring the full parameter space with forward approaches might in general prove infeasible. Moreover, inverse approaches might fail to provide a full picture of the sets of possible density solutions that fit the data within the noise. The problem of non-uniqueness is addressed in \citet{tricarico_global_2013}, where forward methods are employed to explore the full space of exact solutions, as found by determining the null space of the matrix relating the measurements to basis functions of the density expansion. The basis functions are generally orthogonal polynomials, which makes such an approach less suited to model interiors with sharp contrasts between density regions (although a mixed model where additionally the shape of an anomaly is adjusted iteratively as in \citet{silvaInteractiveGravityInversion2006} is already discussed in \citet{tricarico_global_2013}). We focus here on the complementary case of discrete regions of constant density, which might well approximate bodies that have undergone shattering or reaccretion. In particular, we think that our piece-wise constant approach should be well suited to small bodies because their low gravity most likely left their building blocks intact during the accretion process (no material mixing, important density jump, possible large cavities inside).
Other proposed approaches to the gravity inversion of small bodies include multiresolution \citep{sorsa_tomographic_2020} or Markov-chain Monte-Carlo \citep{izquierdoObjectOrientedBayesianGravity2023} bayesian inversion, matrix inversion \citep{ledbetterSmallSatSwarmGravimetry2021}, and machine learning \citep{izzo_geodesy_2022} approaches. In \citet{parkEstimatingSmallBodyGravity2010}, the body was discretized as cubes and a density was estimated for each element, but the solutions were highly affected by the decay of the sensitivity for deeper elements. \citet{takahashi_morphology_2014} combined a linear inversion of the density over a limited number of regions of the body with a forward approach to determine the shape of these regions. Comparison of these methods with our proposed approach will be addressed in Section \ref{sec:Discussion}.  
In recent years there have been multiple real-case applications of gravity inversions for asteroids, starting from the Dawn mission analyses of Vesta \citep{ermakovConstraintsVestaInterior2014}
and Ceres \citep{ermakovConstraintsCeresInternal2017}. The gravity data of OSIRIS-REx for Bennu has been extensively analyzed by \citet{scheeres_heterogeneous_2020}, where the polynomial inversion of \citet{tricarico_global_2013} was combined with a forward approach itself driven by analytical modelling and morphological constraints to derive a possible interior structure of Bennu composed of a light central core and an equatorial ring of lower density than the mantle. From the same dataset, \citet{tricaricoInternalRubbleProperties2021} were able to propose interior models under the assumption of Bennu being a rubble-pile.
As demonstrated by these efforts and due to the non-uniqueness of the problem, inverse methods should still be combined with forward approaches in the exploration of the solution space and possibly in providing realistic uncertainties associated with each solution. \\

%D\slm{\sout{et}ET}AIL (D\slm{\sout{e}E}tail Asteroids Interiors \slm{\sout{t}T}hrough Level-sets \slm{[ALTERNATIVE OPTIONs: AGILE for Asteroid Gravity to Interior  through Level-sEt OR INTEGRAL for INTErior from GRAvity using Level-set or IDEALS for Interior DEnsity mAp using LeveL-Set or GRIDS for GRavity Interior Density level-Set or IMAGES for Interior Map from Gravity Experiment using level-Set\tbd]})
The proposed approach, from hereon referred to as GILA (Gravity Inversion via Level-sets for Asteroids), relies on the use of level-set functions to estimate the shape and location of density anomalies within the body, a technique already well-established in many fields \citep{hedgesStochasticLevelsetMethod2017}, including Earth local-gravity inversion \citep[][and references therein]{giraud_generalization_2021}, but to the best of our knowledge not yet applied to gravity inversion of other bodies.
We strive to keep our approach free from initial assumptions, although a synergy with other methods of interior inference is vital to resolve the inherent degeneracy of the problem. Therefore, while in the current analysis we limit ourselves to the sole use of gravity data and assume no external information about the interior distribution, further efforts will be required to inject into our approach constraints from other observables and different gravity inversions paradigms.

An outline of GILA is presented in Section \ref{sec:methodology}, with Section \ref{sec:simulations} displaying its application in the recovery of a known ground-truth model from simulated observables. Finally, Section \ref{sec:realData} shows how the method fares when applied to real data from the OSIRIS-REx mission. 

% \slm{points that could be mentioned in the introduction
% \begin{itemize}
%     % \item many small bodies have already been visited by spacecraft and many more to come
%     % \item however only half a dozen of these bodies have their gravity field determined behind GM. 
%     % \item We could add a table with Eros, Vesta, Ceres, Chury, Phobos, Bennu (TBD)
%     % \item Previous methods have been proposed in the past (e.g. Tricarico 2013). DESCRIBE them roughly, pointing out their strengths and weaknesses.  BOTH FORWARD AND BACKWARD METHODS SHOULD BE MENTIONED I THINK
%     \item the piece-wise constant approach we propose here is well suited for small bodies because their low gravity would have left their building block intact during accretion process (no material mixing, important density jump, possible large cavities in their inside)
%     % \item Ceres and Vesta are already quite massive, which makes them less suitable for the method in its current version
%     % \item Gravity field of Phobos and Chury was determined to degree-2 only, which makes the inference of the interior density map more difficult, but not impossible as we discuss it later (Sec.XX - I DON'T SEE THE CASE AS A FUCNTION OF GRAV MAX DEGREE EXPANSION - TO BE DISCUSSED)
%     % \item Bennu and Eros are good target to apply our method, and Phobos should be soon too, since MMX should provide a 5x5 field. 
%     % \item ...
%     % \item We first describe the method (Sec.xx), then... 
% \end{itemize}}

\section{Methodology}
\label{sec:methodology}
GILA figures as an extension to small bodies of methods well established in Earth local gravity inversion. We therefore give a high-level overview of the aspects shared with these methods, while focusing on the characteristics implemented to make it suitable to the global gravity inversion of small bodies. 

\subsection{From interior density to gravity field}
The classical expression of the gravitational potential of a generic body makes use of a spherical harmonics expansion \citep[e.g.][]{heiskanen1967physical}:

\begin{equation} \label{eq:kaulaPotential}
U(r, \theta, \phi) = \frac{G\mathcal{M}}{r}\sum_{l=0}^{\infty}\sum_{m=0}^{l}\left(\frac{r_0}{r}\right)^{l} P_{lm}(\cos\theta)\cdot(C_{lm}\cos m\phi + S_{lm} \sin m\phi)
\end{equation} 
with $G$ the universal gravitational constant, $\mathcal{M}$ the mass of the body, $r_0$ the reference radius, $\left( r, \theta,\phi \right)$ the spherical coordinates (radius, colatitude, longitude), $P_{lm}$ the (associated) fully-normalized Legendre polynomials of degree $l$ and order $m$, and $C_{lm}$ and $S_{lm}$ the dimensionless fully-normalized Stokes coefficients. We will use the symbol $\mathcal{O}_{lm}$ to indicate any of $C_{lm}$ or $S_{lm}$ when an expression holds for both sets of coefficients. The $P_{lm}$ functions and $\mathcal{O}_{lm}$ coefficients are obtained from their unnormalized counterparts by respectively multiplying and dividing by the normalization factor $\sqrt{(2-\delta_{m,0})(2l+1)(l-m)!/(l+m)!}$, $\delta_{m,0}$ being the Kronecker delta. In practical applications, the expansion is generally truncated at a given degree $l_{\max}$. 

The Stokes coefficients are typically part of the set of solve-for parameters in an orbit determination campaign, and thus represent the way the gravitational potential is estimated in radio science. As is well known, this expression is well suited for bodies of nearly-spherical shape, but less so for irregular bodies, where the series might display convergence issues when computed within their circumscribing (Brillouin) sphere. For small bodies, alternative representations of the gravity, such as those based on ellipsoidal \citep{parkGravityFieldExpansion2014} or interior \citep{takahashi_small_2014} harmonics, Fast Fourier Transform \citep{perez-molinaFFTGravityField2022a}, or neural networks \citep{izzo_geodesy_2022} might prove more accurate. Nonetheless, spherical harmonics are still usually provided as deliverables of radio science campaigns, for reasons ranging from legacy in orbit determination software to their relative simplicity and direct connection to specific perturbations in the spacecraft orbit \citep{scheeres_geophysical_2016}, and because in most orbit or fly-by configurations the spacecraft does not enter the Brillouin sphere. It is for these reasons that we choose to employ the Stokes coefficients as observables in our study, meaning we assume they are provided as an external input, be it from radio science or from complementary methods.

In turn, the normalized Stokes coefficients are related to the interior density of the body via the integral \citep[e.g.][]{jekeliPotentialTheoryStatic2007}: 

\begin{equation} \label{eq:stokesIntegral}
\begin{bmatrix} C_{lm} \\ S_{lm}\end{bmatrix} = \frac{1}{(2l+1)\mathcal{M}}\int_{\mathcal{V}_\mathcal{B}} \rho {\left(\frac{r}{r_0}\right)}^l\cdot P_{lm}(\cos \theta) \begin{bmatrix} \cos(m\phi) \\ \sin(m\phi)\end{bmatrix}  dV
\end{equation} 
where $\mathcal{V_\mathcal{B}}$ indicates the total volume of the body. Thus, given any expression for the density $\rho$ within the body (e.g. a series of orthogonal functions), the Stokes coefficients can be computed via Eq. \ref{eq:stokesIntegral}. Similarly to \citet{le_maistre_signature_2019}, we discretize Eq. \ref{eq:stokesIntegral} by considering the body as a collection of elements: the volume of the body $\mathcal{V}_\mathcal{B}$ is assumed to be composed of a surface layer and an interior layer. The interior layer is here generally represented by a hexahedral grid (see Figure \ref{fig:interiorVisualization}), but can be any kind of 3D mesh. The surface layer consists of the region between the interior layer and the surface mesh.  We choose this structure to allow for a generic mesh discretization of the interior which is independent of the surface mesh, itself assumed to be an external input and usually composed of triangular elements. The Stokes coefficients of the body are then computed as a weighted sum of the contributions of each element (cells of the interior grid, plus the surface layer), the weight being the element mass:
\begin{equation} \label{eq:stokesDiscrete}
\mathcal{O}_{lm} = \frac{1}{\mathcal{M}} \left[\rho_{s} \mathcal{V}_{s} \mathcal{O}_{lm}^{s} + \sum_{i=1}^{N} \rho(\bm{x}_i) \mathcal{V}_i \mathcal{O}_{lm}^i\right]
\end{equation} 
where $N$ is the total number of grid cells, $\mathcal{V}_i$ the volume of the $i^{th}$ cell, and $\bm{x}_i=(x_i, y_i, z_i)$ the coordinates of its center, while the surface layer is assigned a constant density $\rho_{s}$ and has a volume $\mathcal{V}_{s} = \mathcal{V}_{B} - \sum_{i=1}^{N} \mathcal{V}_i$. The Stokes coefficients of each element, here denoted as $\mathcal{O}_{lm}^s$ and $\mathcal{O}_{lm}^i$, are computed via the line-integral approach of \citet{jametLineIntegralApproach2020}. Compared to other well-established methods for the computation of the Stokes coefficients of a generic polyhedron \citep{werner_spherical_1997, tsoulis_recursive_2009}, this method is non-recursive, allowing for vectorization in the gravity degree-and-order dimensions. While not analytical, it has sufficient accuracy at the low degrees we consider in our study even with a coarse discretization of the line integrals (we generally use $2^3$ line integration points for each edge of the exterior mesh and the internal grid). Moreover, it is not restricted to polyhedra with a vertex at the origin of the potential expansions, allowing for a more immediate modelling of heterogeneities within the body, compared for example to the approach of \citet{takahashi_morphology_2014}. Overall, the computational overhead of this polyhedron model implementation compared to the simpler mascons approach \citep[e.g.][]{tardivel_limits_2016} is not impractical, especially when high accuracies, and therefore more mascons, are required.

Eq. \ref{eq:stokesDiscrete} provides a linear relation between the gravity coefficients and the density of each grid cell, itself computed as the value of the density function at the location of the cell center ($\bm{x}_i$), which corresponds to a node in the discretization grid. The density distribution within the body is here assumed to be piecewise-constant, meaning that the elements are grouped to form density anomalies (regions of uniform density). Hence:
\begin{equation} \label{eq:rhoPiecewise}
\rho(\bm{x}) = \sum_{j=0}^{M} \rho_j \mathds{1}_j(\bm{x})
\end{equation} 
where $M$ is the total number of density anomalies and $\mathds{1}_j$ is the indicator function of the $j^{th}$ anomaly, which is 1 inside the anomaly and 0 outside. For better clarity, we will reserve the index $j$ to describe any of the $M$ anomalies, and the index $i$ to refer to any of the $N$ grid elements. The index $j=0$ refers to the background density $\rho_0$, for which the indicator function is $\mathds{1}_0=1$ over the whole body, and which is also the density of the surface layer (that is, we impose $\rho_s = \rho_0$). The parameter $\rho_j$ then represents the density jump with respect to $\rho_0$, associated with the $j^{th}$ anomaly. This way, heterogeneous interior models can be constructed by defining $M$ domains for the density anomalies and assigning the corresponding density jumps. As in \citet{li_multiple_2017} and \citet{giraud_generalization_2021}, we don't allow for overlap of anomalies, meaning that at each point inside the body only one of the $\mathds{1}_j$ apart from $\mathds{1}_0$ can have a value of 1. This is handled in our algorithm on a "last-in-first-out" basis, meaning that in case of overlapping of anomalies the $\mathds{1}_j$ of highest $j$ index is assigned a non-zero value, while all the other indicator functions (save for $\mathds{1}_0$) are set to 0.

%Contrary to \citet{li_multiple_2017}, where each point in space belongs to a single anomaly and to the background region, in our nominal settings we do not envision special treatment for regions where multiple anomalies intersect. In those regions, the total density jump will be sum of the jumps from all contributing anomalies, further expanding the range of models that can be represented. An alternative approach based on \citet{giraud_generalization_2021}, avoiding overlapping of the anomalies, was also explored. \slm{SAY MORE... what are the outcomes of such an 'exploration' in short?}

\subsection{Density function estimation}
The linear relation of Eq. \ref{eq:stokesDiscrete} forms an under-determined system that can be inverted to obtain a vector of local densities $\bm{{\rho}} = \left(\rho_s, \rho(\bm{x}_1),...,\rho(\bm{x}_N)\right)$ over all the discretization elements. 
However, doing so without any regularization terms or depth-weighting of the partials will give a density solution with anomalies concentrated at the surface of the body, where the strength of the partials is larger \citep{parkEstimatingSmallBodyGravity2010}. Depth-weighting in such an underconstrained case was explored in several Earth local-gravity inversion studies \citep[e.g.][]{li_3-d_1998}, but our attempts to extend it to the global gravity inversion of small bodies were so far unsuccessful.
On the other hand, the regularizing assumption of piecewise-constant density (Eq. \ref{eq:rhoPiecewise}) allows to substantially reduce the number of parameters to estimate, at the cost of a limitation in the range of possible models. With these assumptions, the $M$ density jumps of each anomaly can be estimated, along with the background density, in place of the $N+1$ densities of each discretization element (where $M<<N$), as in \citet{takahashi_morphology_2014} for instance. This linear estimation problem is generally over-determined, as long as the number of zones $M$ is selected to be below $(l_{\max}+1)^2$ \citep{scheeres_estimating_2000}, and can be solved via weighted least-squares. However, since the indicator functions $\mathds{1}_j$ are fixed in the estimation, the solution will strongly depend on their value, and therefore on the initial subdivision of the body. %GILA addresses this problem by adjusting both the densities and shapes of the anomalies.

\subsubsection{Level-set method}
\begin{figure}[!htbp]
      \begin{center}
        \includegraphics[width=\textwidth]{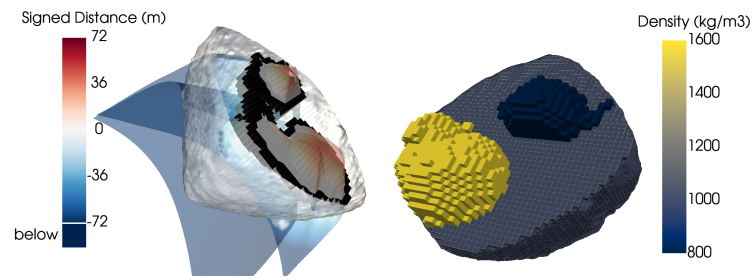}
      \end{center}
      \caption{Level-set representation of the density. Left: values along the $Y=0$ plane of 2 level-set functions ($\phi_1$ and $\phi_2$), obtained as the signed-distance function of the Stanford bunny and the Utah teapot. The black cubes indicate the narrow-band region. Right: density model produced by these 2 level-set functions, assuming as density contrasts $\rho_1 = \qty{600}{kg/m^3}$ and $\rho_2 = \qty{-200}{kg/m^3}$ , and as background density $\rho_0 = \qty{1000}{kg/m^3}$ .}
      \label{fig:interiorVisualization}
\end{figure}

In order to reduce the influence of the initial zones subdivision on the density solution, the shape of the anomalies is here estimated along with their density contrast. The shape derivative is computed via the level-set method, as in \citet{giraud_generalization_2021}. The boundary of each anomaly is defined implicitly as the 0 level-set of a scalar function $\phi : \mathbb{R}^3 \mapsto \mathbb{R}$ (Figure \ref{fig:interiorVisualization}). Clearly there are infinite level-set functions $\phi$ providing the same anomaly boundary, but commonly $\phi$ is taken to be the signed-distance function from the boundary, in order to improve the numerical stability of iterative methods \citep{osherLevelSetMethods2003}. We compute the signed-distance via the \emph{scikit-fmm} \footnote{\url{https://github.com/scikit-fmm/scikit-fmm}} Python implementation of the fast-marching method \citep{sethianFastMarchingLevel1996}.

The indicator function of the $j^{th}$ anomaly is then defined as:
\begin{equation} \label{eq:indicatorHeaviside}
\mathds{1}_j(\bm{{x}}) = H(\phi_j(\bm{{x}}))
\end{equation} 
where $\phi_j(\bm{x})$ is the level-set function specific to the $j^{th}$ anomaly, and in our case $\bm{x}$ takes the discrete values of the centers of the grid elements. $H$ is the Heaviside (or step) function, so that:
\begin{equation} \label{eq:heavisideCases}
H(\phi_j) = \begin{cases}
  1  & \text{if } \phi_j \geq 0 \text{  (inside the anomaly)}\\
  0 & \text{otherwise (outside the anomaly)} 
\end{cases}
\end{equation} 

By combining Eqs. \ref{eq:stokesDiscrete}, \ref{eq:rhoPiecewise}, and \ref{eq:indicatorHeaviside}, we therefore obtain an expression of the Stokes coefficients in terms of the density jump of each anomaly and of the values of the corresponding level-set function at the grid nodes:
\begin{equation} \label{eq:stokesLevelSet}
\mathcal{O}_{lm} = \frac{1}{\mathcal{M}} \left[\rho_s \mathcal{V}_s \mathcal{O}_{lm}^s +\sum_{i=1}^{N}\sum_{j=0}^{M} \rho_j H(\phi_j^i) \mathcal{V}_i \mathcal{O}_{lm}^i\right]
\end{equation} 
where, for ease of notation, $\phi_j^i$ denotes $\phi_j(\bm{x}_i)$, that is the value of the $j^{th}$ level-set function at the location of the $i^{th}$ grid element. In order to generalize our conclusions to bodies of any shape and size, we normalize the signed-distance function, dividing it by the minimum grid resolution. \\
For practical applications where an analytical derivative is advantageous, a differentiable approximation of the Heaviside function is generally preferred, such as:
\begin{equation} \label{eq:heavisideContinuousCases}
H(\phi_j) = \begin{cases}
  1  & \text{if } \phi_j > \tau \\
    \frac{1}{2}+\frac{\phi_j}{2\tau} + \frac{1}{2\tau}\sin{\left(\frac{\pi \phi_j}{\tau}\right)}  & \text{if } \left|\phi_j \right| \leq \tau\\ 
  0  & \text{if } \phi_j < -\tau
\end{cases}
\end{equation} 
where $\tau$ is a distance defining the maximum distance from the anomaly boundary where the level-set function is allowed to vary (black cubes in Figure \ref{fig:interiorVisualization}). This process is known as \emph{narrow-banding} \citep{adalsteinssonFastLevelSet1995}. The parameter $\tau$ is usually taken to be a small multiple of the grid resolution, expressed as the minimum grid element size $\Delta \bm{x}$: here nominally $\tau = 1.5 \min(\Delta x, \Delta y, \Delta z)$.
From the continuous approximation of Eq. \ref{eq:heavisideContinuousCases}, it is easy to compute the partial derivatives of the anomaly indicator function with respect to the values of the level-set function at the grid nodes. However, this approximation proved to be less stable in our iterative optimization, 
possibly because of vanishing partials for cells near the edge of the narrow-band increasing the condition number of the observation matrix. For this reason, while still opting for the narrow-banding philosophy due to its computational efficiency, we use the non-differentiable expression for the Heaviside function (Eq. \ref{eq:heavisideCases}), with approximate partials coming from the signed-distance property of $\phi_j$:
\begin{equation} \label{eq:numericalPartials}
\frac{\partial \mathds{1}_j}{\partial \phi_j} = \frac{1}{\left|\phi_j\right|+\xi}
\end{equation} 
$\xi$ being a small positive term that limits the value of the partials when $\phi_j \mapsto 0$.
% This is equivalent to approximating the Heaviside function by the smeared expression of Eq.\ref{eq:heavisideContinuousCases} for the points at the border, with $\tau=\mu$, and propagating the partials according to their distance from the border as in \citet{osherLevelSetMethods2003}.
This expression reflects the fact that if the value of $\phi_j^i$ changes by $-\phi_j^i$, the border of the $j^{th}$ anomaly shifts to the center of the $i^{th}$ cell, meaning that $\mathds{1}_j^i$ varies by $\pm 1$. Small values of $\xi$ provide a better approximation of the numerical partials, but on the other hand lead to high condition numbers of the observation matrix. Here we take $\xi=0.1$, which balances these two effects. 

The partials of the Stokes coefficients with respect to the level-set values of each anomaly and at the location of each grid element are then computed via the chain rule:

\begin{equation} \label{eq:cnmPartialsPhi}
\frac{\partial \mathcal{O}_{lm}}{\partial \phi_j} = \frac{\partial \mathcal{O}_{lm}}{\partial \rho} \rho_j \frac{\partial \mathds{1}_j}{\partial \phi_j} 
\end{equation} 
%\label{eq:derparLVLset}
where, from Eq. \ref{eq:stokesDiscrete} and as in \citet{scheeres_estimating_2000}:
\begin{equation} \label{eq:cnmPartialsRho}
\frac{\partial \mathcal{O}_{lm}}{\partial \rho} = \frac{1}{\mathcal{M}} \sum_{i=1}^{N_j} \mathcal{V}_i \mathcal{O}_{lm}^i
\end{equation} 
with $N_j$ the number of grid elements contained in the $j^{th}$ anomaly.

\subsubsection{Iterative least-squares solution}
\label{sec:leastSquaresTheory}
Estimating the shape of the anomaly along with its density jump leads to a non-linear problem. The target parameters for each anomaly are its density jump $\rho_j$ and the nodal values of the associated level-set function, i.e. the set $\{\phi_j^i\}_{i=1,...,N}$, although in our case $i$ only takes the index-values of the grid elements within the narrow-band of the anomaly. We minimize the residuals between the observed Stokes coefficients and those computed from Eq. \ref{eq:stokesLevelSet} in a least-squares sense. Given the partial derivatives defined above, equation \ref{eq:stokesLevelSet} is linearized in a neighborhood of the current values of the target parameters. If $\bm{y}$ is the vector of $P\leq(l_{\max}+1)^2$ residuals and $\bm{t}=\left(\rho_0,...,\rho_M, \phi^1_1, ..., \phi^N_1, \phi^1_2, ..., \phi^N_M\right)$ the vector of $K$ solve-for parameters, then:
\begin{equation} \label{eq:linearizedProblem} 
\bm{y} = J \bm{\delta t} +\bm{\varepsilon}
\end{equation}
$\bm{\delta t}$ being a small correction to the parameters, and $J$ the $P\times K$ Jacobian matrix, where the term $J_{pk}$ is computed from Eq. \ref{eq:cnmPartialsRho} or Eq. \ref{eq:cnmPartialsPhi} based on whether the $k^{th}$ parameter is a density or a level-set correction, respectively. The measurements errors $\bm{\varepsilon}$ are generally correlated, but the associated covariance matrix $\Omega \in \mathbb{R}^{P\times P}$ on the Stokes coefficients can be assumed to be known, as it is an output of a radio-science campaign. These correlations can be accounted for in a generalized least-squares approach \citep[e.g.][]{bjorckNumericalMethodsMatrix2015}: if $\Omega$ is assumed to be symmetric positive definite, it has a unique Cholesky factorization $\Omega = WW^T$, with $W\in \mathbb{R}^{P\times P}$ nonsingular. By pre-multiplying both sides of Eq. \ref{eq:linearizedProblem} by $W^{-1}$, we then transform the system so that the elements on the left-hand side have associated errors which are uncorrelated and have the same variance. The whitened residuals vector is then $\bm{y'} = W^{-1}\bm{y}$ and the corresponding observation matrix $J' = W^{-1}J$. In the following we drop the prime symbol, so that $\bm{y}$ indicates residuals with uncorrelated and homoscedastic errors and $J$ their observation matrix. In the simulation tests of Section \ref{sec:simulations} we assume no correlations in the synthetic $\mathcal{O}_{lm}$, so that the problem is reduced to a weighted least-squares, where $W = \diag(\sigma_1,...,\sigma_P)$ is a diagonal matrix with the standard deviations of the measurement errors as elements.\\

The objective function ($\Psi$) we wish to minimize is then that of ordinary least-squares, with additional penalty terms:
\begin{equation}
\label{eq:levelSetObjective}
\Psi(\bm{\delta t}) = {\lVert \bm{y}-J\bm{\delta t} \rVert}_2^2+\lambda^2{\lVert\bm{\delta t} \rVert}_2^2+\Psi_c(\bm{\delta t})
\end{equation}
where ${\lVert \cdot \rVert}_2$ is the 2-norm. The second term in the expression of $\Psi$ is the Tikhonov regularization term \citep[e.g.][]{bjorckNumericalMethodsMatrix2015}, which penalizes large corrections on the parameters. It is necessary in order to render the method robust with respect to noise in the data, and its weight $\lambda$ can be assigned empirically, or via generalized cross-validation (GCV). The last penalty term is a generic quadratic constraint, where available. For example, one could search for solutions close to a given reference (\emph{a priori}) density distribution. Then, if $\bm{t'}$ are the values of the solve-for parameters in the reference model, and $\bm{t}$ their values at the current iteration, the penalty term could be:
\begin{equation}
\label{eq:levelSetConstraints}
\Psi_c(\bm{\delta t}) = \nu^2{\lVert \bm{t'}-\bm{t}-\bm{\delta t} \rVert}_2^2
\end{equation}
where $\nu$ is the weight of the constraint, which in this case favours corrections bringing the current parameter values closer to those of the reference model. In practice, these quadratic penalties can be treated as supplemental observations and added as extra rows to Eq. \ref{eq:linearizedProblem}.
We use a Gauss-Newton method to iteratively adjust the target parameters, so that at each iteration the correction $\bm{\delta t}$ estimated from the linearized system is simply added to the vector of parameters. The system is inverted using the SciPy implementation of the LSQR algorithm \footnote{\url{https://docs.scipy.org/doc/scipy/reference/generated/scipy.sparse.linalg.lsqr.html}}, which can handle the sparse matrices resulting from the addition of penalties. Alternatively, the \emph{RidgeCV} function of the {scikit-learn} Python library \citep{scikit-learn} is used to solve the Tikhonov-regularized problem by means of GCV.\\

More often than not, the objective function turns out to be non-convex. Similarly to restarts in gradient-descent optimization, we mitigate the problem of early convergence to local minima by periodically amplifying the estimated $\bm{\delta t}$ so that the maximum level-set correction is larger than the narrow-band parameter $\tau$. For the same reason, and because we assume no prior knowledge about the interior of the body, we nominally start from initial models where the different (usually 3 or more) level-set functions take positive values all over the body.
% As mentioned above, the prior model can alternatively obtained from the averaging approach, but this may lead more easily to early convergence if the average model is concentrated in regions not including the target anomalies. \slm{(SAY MORE, e.g. Although, not affecting much our results mainly AND YOU CAN ADD WHY IF RELEVANT, "because...")}.

\begin{figure}[!htbp]
      \begin{center}
        \includegraphics[width=0.5\textwidth]{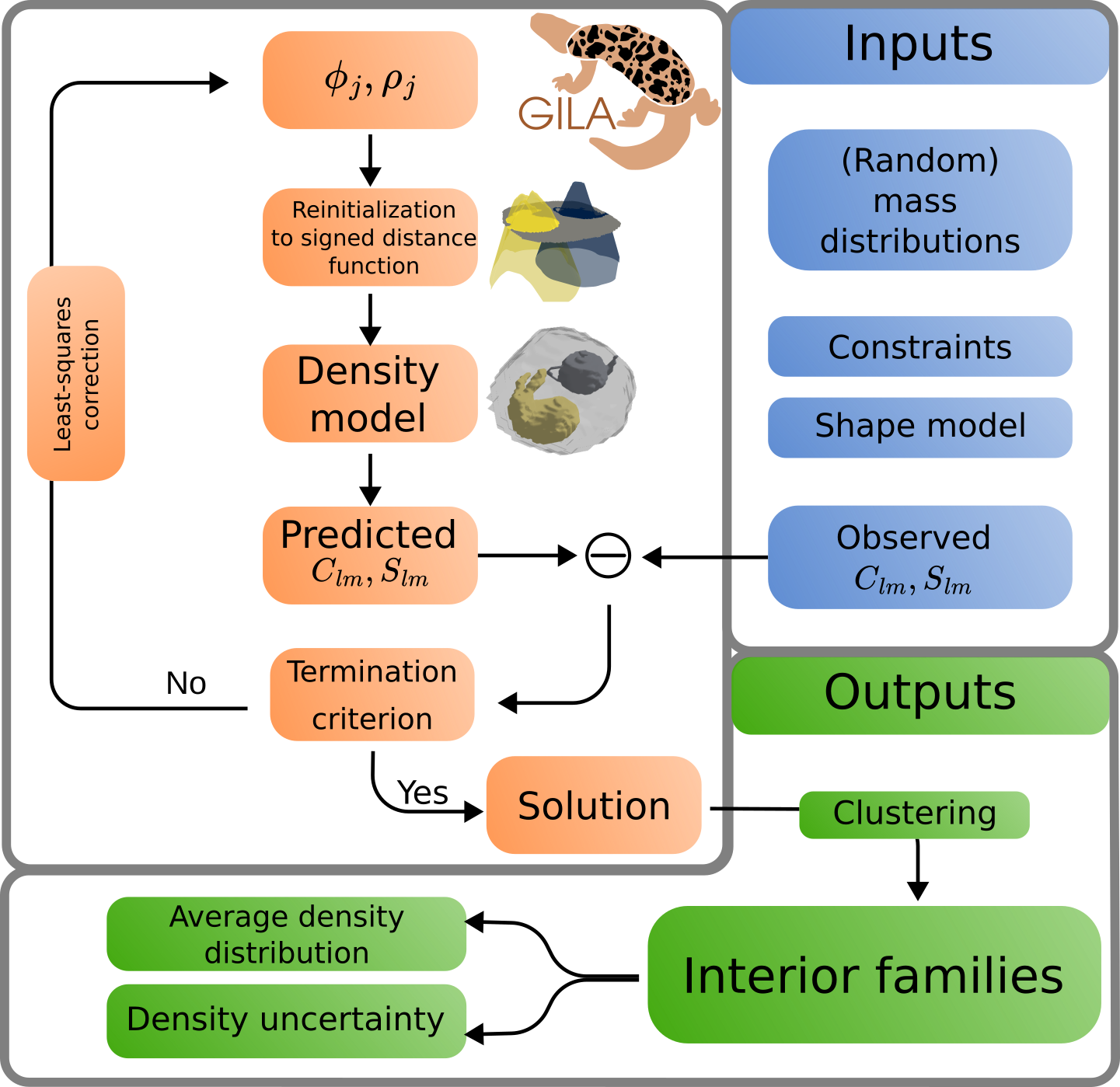}
      \end{center}
      \caption{Pipeline of the inversion algorithm}
      \label{fig:pipeline}
\end{figure}

Figure \ref{fig:pipeline} shows a schematic of GILA. The orange section (left-hand side of the figure) is the core part of the method and consists of the iterative least-squares inversion, which provides a single density solution given the inputs. The latter are represented in blue, in the top-right region, and include: the observables, namely the Stokes coefficients $\mathcal{O}_{lm}$; the shape model of the body, in the form of a triangular mesh; eventual quadratic constraints, which would help restrict the range of possible solutions; a set of initial density distributions, as running the inversion algorithm with different \emph{a priori} is advised given the non-convexity of the problem. For the moment, GILA does not support running multiple inversion loops in parallel. A single run of the iterative least-squares until convergence requires on average 3 minutes on a laptop with CPU Intel i7-12700H (14 cores, up to 4.70 GHz) for an interior grid of 50 elements per dimension, while the time for the computation of the $\mathcal{O}_{lm}$ (which are stored once computed) depends on the resolution of the mesh and the degree of the gravity, but is itself at most a few minutes in the cases considered here. If several initial models are used, the outputs will be a set of solutions which may fall into different regions of the solution space. In order to distinguish between mass distributions which agree with the data within the noise but are widely different from each other, we use global quantities independent from the input measurements, such as the moments of $\rho(\bm{x})$ (see Eq. \ref{eq:densityMoments}). Then, as displayed in the green section of the figure, the set of solutions can be subdivided into a small number of interior families via a clustering algorithm, and each family can be characterized on the basis of its average distribution and the spread of the density values at each grid element.

\section{Synthetic retrievals}
\label{sec:simulations}
In this section, we assess the ability of the method to retrieve a known density distribution from synthetic gravity measurements. In each case, the known density model is used for the computation of synthetic observed Stokes coefficients $\mathcal{O}_{lm}$ according to Eq. \ref{eq:stokesDiscrete}. Then, the iterative least-squares algorithm is run until either the reduced chi-squared of the residuals (as defined in Section \ref{sec:chiSquared}) reaches a minimum threshold of 0.1 and the model has stopped evolving (after a warm-up of 500 iterations), or a set limit of maximum iterations is reached. Generally, for the first 100 iterations the density contrasts are fixed, and only the background density and the level-set parameters (meaning the shapes of the anomalies) are adjusted. This appears to help avoid early convergence of the solution, although it makes the iterations more dependent on the initial contrasts assigned to each density anomaly. \\
We start with noise-free measurements, in order to gauge how the method is robust with respect to the non-uniqueness of the solution and the non-convexity of the objective function. We then present more realistic scenarios where the synthetic data are contaminated with Gaussian noise, according to different noise profiles, with the target of testing the behaviour of the method with respect to another characteristic of ill-posed problems, namely their instability.

\begin{figure}[!htbp]
      \begin{center}
        \includegraphics[width=\textwidth]{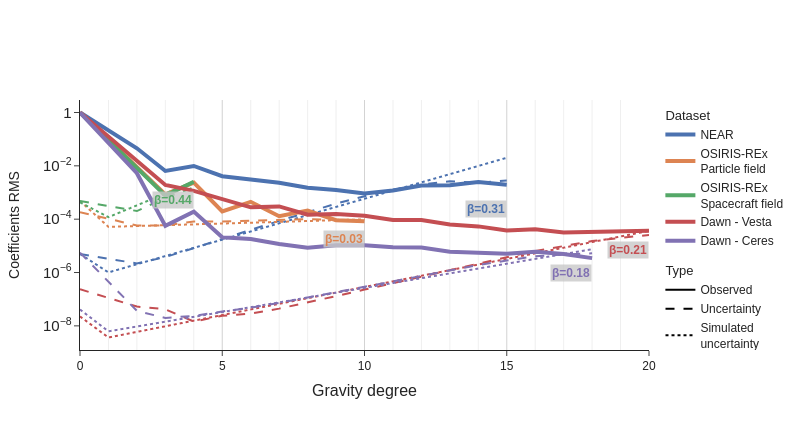}
      \end{center}
      \caption{Power spectrum of measured coefficients (solid lines) with corresponding uncertainties (dashed lines) for the Eros gravity field of NEAR \citep{miller_determination_2002-1}, the two OSIRIS-REx gravity solutions for Bennu \citep{scheeres_heterogeneous_2020}, and the two Dawn gravity solutions for Vesta \citep{konopliv_vesta_2014} and Ceres \citep{konopliv_ceres_2018}. The dotted lines represent the noise profile obtained from Eq. \ref{eq:noiseSpectrum}, assuming $l_{\max}$ to be right before the eventual intersection between the solid and dashed lines. The values of $\beta$ are computed as the average slope of the dashed lines over the range of degrees where they show log-linearity.}
      \label{fig:noiseProfilePlot}
\end{figure}

The absolute uncertainty of the Stokes coefficients measured in a radio science campaign is usually higher for high degrees, since their lower signal-to-noise ratio means their effects on the noisy trajectory reconstruction are harder to decorrelate than for low degrees. This could however not always be a fair prediction. For example, when the coefficients are obtained via optical tracking of orbiting dust particles they might display a flat uncertainty profile as a function of the degree \citep{chesleyTrajectoryEstimationParticles2020}.
Still, the standard deviations of the synthetic measurements errors are here assigned following a noise profile varying as a function of the harmonic degree and based on the exponential expression of \citet{tricarico_global_2013}, itself derived from the uncertainty spectrum of the gravity coefficients estimated for Eros using NEAR radio-tracking data \citep{miller_determination_2002-1}.
The noise spectrum is then computed as:
\begin{equation} \label{eq:noiseSpectrum}
    \sigma(l) = \alpha 10^{\beta(l-l_{\max})} \sqrt{\frac{\mathcal{S}_{l_{\max}}}{2 l_{\max} + 1}},
\end{equation}
where $l_{\max}$ is the maximum harmonic degree of the measured (or simulated) Stokes coefficients and $\mathcal{S}_{l_{\max}}$ is their power spectrum (sum of their squares) at the degree $l_{\max}$. 
The parameter $\alpha$ indicates instead the relative strength of $\sigma(l_{\max})$ compared to $\mathcal{S}_{l_{\max}}$. We therefore talk of $100\%$ noise to indicate a case where 
$\alpha=1$, meaning that most of the coefficients of degree $l_{\max}$ are consistent with 0. Similarly, 50\%,10\%, 1\% noise values correspond to $\alpha$ values of 0.5, 0.1, and 0.01, respectively.
The coefficient $\beta$ is the slope parameter, indicating how fast the uncertainty decreases for degrees lower than $l_{\max}$. Figure \ref{fig:noiseProfilePlot} shows the power spectrum of measured coefficients and their uncertainties for 5 different real datasets, as well as the $\beta$ value approximating their noise profile. Degree-1 values and uncertainties are not provided in the real datasets because assumed to be known and equal to 0, meaning that the potential is expressed in a frame with its origin at the center of mass of the body. The uncertainties resulting from Eq. \ref{eq:noiseSpectrum} tend to underestimate the GM uncertainty, which is why we multiply the degree-0 $\sigma$ by a factor of 10. The simulated noise profile also underestimates the errors of the Dawn datasets at low degrees. As in \citet{tricarico_global_2013}, in all our simulations we assume $\beta=1/3$, which is roughly the value fitting the NEAR measured uncertainties, and sits between those of the OSIRIS-REx spacecraft-field uncertainties and those of the Dawn datasets.\\

\subsection{Inversion quality metrics}
\label{sec:qualityMetrics}
The estimated density model is evaluated in terms of its accuracy in both fitting the synthetic measurements within the given noise and reproducing the ground-truth density distribution. Specifically, we test two different quality metrics of the inversion: the reduced $\chi^2$ statistic of the whitened measurements residuals and the Pearson correlation between the estimated and the true density distributions. We look at the correlation because it appears to be less model-dependent compared to other measures of the similarity between interior models, such as the RMS of the density errors at each grid element with respect to the ground-truth, or the relative error on the total mass of the anomalies. However, of all these metrics only the reduced $\chi^2$ is applicable to real data analysis, and while it can provide an indication of good fit, the corresponding solution is not guaranteed to correctly represent the true interior of the body.

\subsubsection{Reduced $\chi^2$}
\label{sec:chiSquared}
The reduced $\chi^2$ statistic, denoted as $\chi_P^2$, is computed as the ratio of the weighted residuals sum-of-squares (RSS) and the number of measurements (P), that is:
\begin{equation}
    \chi_P^2 = \frac{\sum_p{y_p^2}}{P}
\end{equation}
Values of $\chi_P^2$ close to 1 indicate a good fit consistent with the given measurement weights, while $\chi_P^2>>1$ suggests an inaccurate fit and $\chi_P^2<<1$ an overfitting of the data (where the model also fits the noise). From simulations we observed that the latter case, that of overfitting, is very rare for GILA, due to the values of the objective functions being limited by the model error. This error is generally related to the inability of the inversion model to reproduce exactly the ground-truth with its parameters, but in our case an even larger contribution is due to the fact that the iterative inversion algorithm is rarely able to reach exactly the global minimum, but instead convergence happens at best in a neighborhood of the optimum. This is justified by the non-convexity of the objective and possibly by the partials of Equation \ref{eq:numericalPartials} approximating less well the gradient of the objective function around a minimum. \\

\begin{figure}[!htbp]
      \begin{center}
        \includegraphics[width=\textwidth]{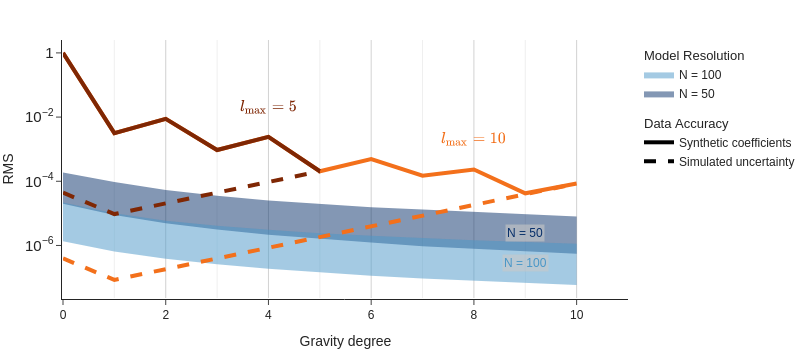}
      \end{center}
      \caption{Comparison between model resolution and measurements accuracy for a Bennu-shaped body. The blue shaded areas are envelopes of the model resolutions for 100 random density distributions, using a grid of 50 (darker area) or 100 (lighter area) elements per dimension. Solid and dashed orange lines indicate the RMS of the synthetic coefficients and corresponding errors, respectively, for $l_{\max}=5$ (darker lines) and $l_{\max}=10$ (lighter lines). The measurements values are computed for a homogeneous distribution, and their uncertainties follow the noise profile of Equation \ref{eq:noiseSpectrum} with $\alpha=1$ (i.e. in the 100\% noise case) and $\beta=1/3$.}
      \label{fig:modelError}
\end{figure}

In order to predict the effect of this inaccurate convergence on the final residuals, we introduce the concept of model resolution.
The grid discretization of the interior implies a non-dense image of the Jacobian matrix $J$, since the smallest possible variation in the density model (e.g. expanding an anomaly by a single grid element) results in a finite variation in the predicted measurements. We characterize this model resolution by looking at the columns of $J$ corresponding to the level-set parameters. For each row of the matrix, the maximum of its absolute values over these columns represents the minimum resolution of the model for the corresponding measurement. This is because the level-set partials in $J$ are the measurements variations per unitary change of the level-set function, and given our normalization of $\phi_j$ by the grid-size, this unitary change is a change of 1 grid element. Trivially, the model resolution thus computed decreases with the density contrast of the estimated anomalies (see Eq. \ref{eq:cnmPartialsPhi}). It also decreases with the decrease of the grid elements' size, itself tied however to higher computation times and a higher degeneracy of the problem. Figure \ref{fig:modelError} shows that with 50 grid elements per dimension, the model resolution is higher in magnitude than the uncertainty of the measurements for low degrees, and drops quickly for higher degrees, while the noise profile increases. The same behavior is observed for 100 grid elements per dimension, but this time the intersection between the model resolution and the data noise happens at a lower gravity degree. This trend and the gravity degrees at which the model resolution intersects the synthetic uncertainties are similar for bodies of different shapes and sizes, although the absolute values of the model resolutions are body-dependent.\\
In simulations, we often test GILA on noise profiles where for lower degrees the measurements $\sigma$ might be smaller than the model resolution. Therefore, even converged models which differ by as little as 1 grid element from the truth might show large values of $\chi_P^2$. To better assess the convergence of the solution, we replace the measurements weights with the model resolution for data points where the latter is larger than the input measurements uncertainty. Therefore, the reduced $\chi^2$ statistic becomes a measure of the agreement between the predicted and the observed measurements when both the data uncertainty and the model resolution are taken into account. Of course, for real data applications, the $\chi_P^2$ with only the data uncertainty should also be considered. Until further improvements to GILA, if the residuals are limited by the model resolution the options are to increase the grid resolution (where feasible, see effect of doubling the resolution in Figure~\ref{fig:modelError}) or to accept a data fit limited by small errors in the model.

\subsubsection{Density correlation}
The Pearson correlation between the ground-truth and retrieved density distribution is computed as \citep[e.g.][]{leerodgersThirteenWaysLook1988}:
\begin{equation}
    \rho_{CORR} = \frac{\sum_i{\left(\rho_i^{(est)}-\overline{\rho}^{(est)}\right)\left(\rho_i^{(sim)}-\overline{\rho}^{(sim)}\right)}}{\sqrt{\sum_i{{\left(\rho_i^{(est)}-\overline{\rho}^{(est)}\right)}^2}
    \sum_i{{\left(\rho_i^{(sim)}-\overline{\rho}^{(sim)}\right)}^2}}}
\end{equation}
where, $\bm{\rho}^{(est)}$ is the vector of estimated density values over each element grid, and $\bm{\rho}^{(sim)}$ the ground-truth density values interpolated over the inversion grid, while $\overline{\rho}^{(est)}$ and $\overline{\rho}^{(sim)}$ are their mean values.

%\subsubsection{RMS density error}
%This metric is simply computed as:
%\begin{equation}
%    \rho_{RMS} = \sqrt{\frac{1}{N}\sum_{i=1}^{N}{{\left(\rho_i^{(est)}-\rho_i^{(sim)}\right)}^2}}    
%\end{equation}
%where $\rho_i^{(est)}$ is the value of the density function at the $i^{th}$ element of the interior grid, and $\rho_i^{(sim)}$ is the value of the density function used in simulation \slm{(i.e. to generate the synthetic data)} and linearly interpolated at the location of the center of the $i^{th}$ element of the estimation grid, since the simulation and estimation grids are not necessarily the same.
%
%\subsubsection{Relative mass error}
%This quantity measures the relative error in the total mass of the estimated anomalies compared to the total mass of the anomalies in the ground-truth model. We indicate with $\Delta \mathcal{M}$ the total mass of the anomalies, given by $\sum_{j=1}^M\left(\rho_j\sum_{i=1}^{N_j}V_i\right)$. Then, the metric is defined for $\Delta \mathcal{M}^{(sim)}\neq 0$ as:
%\begin{equation}
%    \Delta \mathcal{M}_{err} = \frac{\Delta \mathcal{M}^{(est)}-\Delta \mathcal{M}^{(sim)}}{\Delta \mathcal{M}^{(sim)}}
%\end{equation}

\subsection{Target models}
\label{sec:targetModels}
We mostly limit our assessment of the method to three reference models, denoted with numbers 1,2,3 with increasing level of complexity. Specifically: Model 1 is composed of a single off-centered positive anomaly of regular shape, namely a triangular prism, with a density contrast of \qty{600}{kg/m^3}; Model 2 has a single positive anomaly of \qty{600}{kg/m^3} centered at the origin and of complex shape, based on the Stanford dragon; Model 3 has multiple anomalies of complex shapes and different density contrast, in both sign and magnitude (2 positive ones of \qty{600}{kg/m^3}, and 2 negative ones of \qty{-200}{kg/m^3}, and \qty{-600}{kg/m^3}). All three models are piecewise-constant but approximating a wide range of scenarios where the GILA could be applied. Degenerate cases, such as those with concentric shells, are not discussed here as their severe non-uniqueness hinders any possible conclusions about the performance of GILA. \\

\begin{figure}[!htbp]
      \begin{center}
        \includegraphics[width=0.7\textwidth]{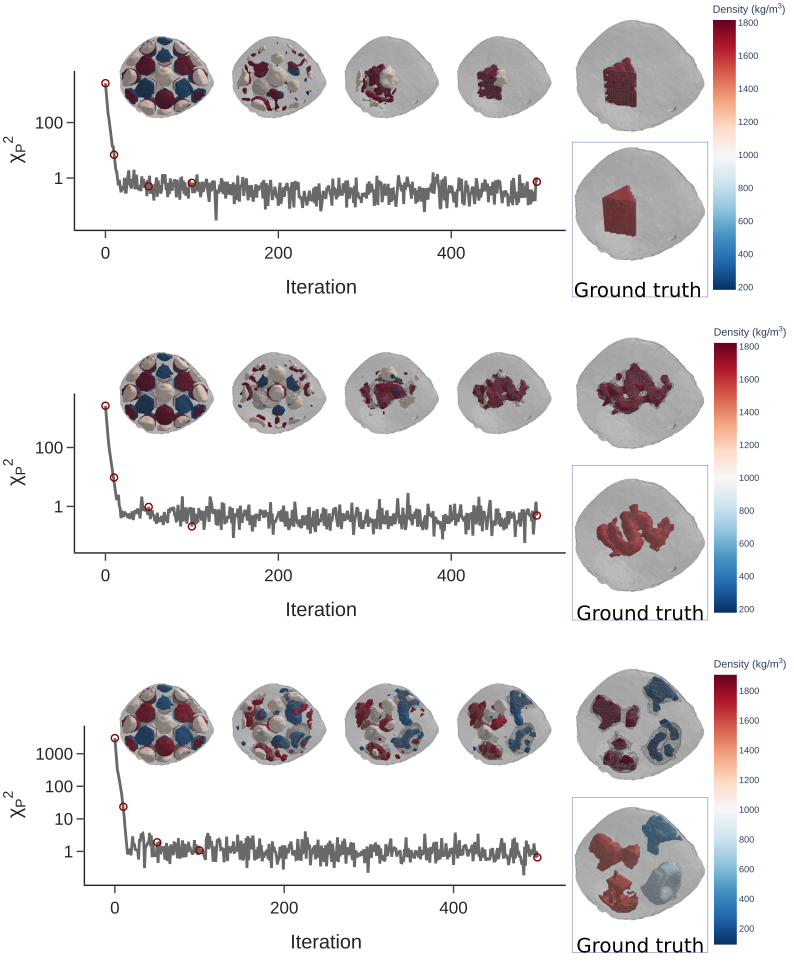}
      \end{center}
      \caption{Three synthetic retrievals of the density models at the bottom right of each subpanel, corresponding to Models 1, 2, and 3, from top to bottom. The line plots represent the values of $\chi_P^2$ over the least-squares iterations, with red circles indicating specific steps of the convergence displayed in the 3D plots, in the same order. The right-most point and its related 3D plot correspond to the solution at convergence. The 3D plots are obtained via the marching-cubes algorithm (as implemented in the scikit-image Python library). The density colorscale is shared between all 3D plots of a given subpanel. The 3D meshes employed for reference Model 3 are Spot and Bob (courtesy of Prof. Keenan Crane), \#3DBenchy (\url{http://3dbenchy.com/}), and Suzanne (Blender Foundation). }
      \label{fig:referenceModels}
\end{figure}

Figure \ref{fig:referenceModels} shows the three reference models and the corresponding retrieved models at convergence when fitting synthetic Stokes coefficients up to degree 11 for a Bennu-shaped body. These plots are meant to show the performance of the model in a purely ideal case, assuming a 1\% noise profile for the measurements weights but no noise perturbation on the measurements and consequently no regularization term in the objective function, which is only composed of the data misfit. Moreover, the shape model used in these simulations is the same as that of the inversions. As density and shape parameters are highly correlated, even at convergence the detected anomalies have density jumps which are not exactly those of the true model, while still agreeing with the data at the noise level, as suggested by the value of the metric $\chi_P^2\sim1$. Nevertheless, Figure \ref{fig:referenceModels} proves that GILA is able to retrieve number, sign, and approximate shape of multiple density anomalies without injection of any prior knowledge about the ground truth. This convergence to the true model, while seemingly clashing with the non-uniqueness of the true inversion, is justified by the reference models being composed of isolated anomalies with relatively high density contrast, by the high resolution of the gravity measurements, and by the forward and inverse discretization grids being identical. 

\subsection{Robustness}
In this section, we build towards an assessment of the GILA over more realistic inversion scenarios, and at the same time attempt to delineate the limits of applicability of the method. To this end, we modify one or more of the scenarios of Figure \ref{fig:referenceModels} by one property at a time, and check how this impacts the inversion quality metrics. The inversion settings varied here are: the degradation of the synthetic measurements, both in their resolution (Section \ref{sec:vsGravResolution}) and level of noise contamination (Section \ref{sec:vsNoise}); the shape and size of the body (Section \ref{sec:vsBodyShape}); the error on the body shape (Section \ref{sec:vsShapeError}); the resolution of the internal discretization grid (Section \ref{sec:vsInteriorGrid}); the size, depth, and density contrast of the anomaly (Section \ref{sec:vsAnomalyDepth}); the initial density model in the iterative fit (Section \ref{sec:vsInitialModel}).

\subsubsection{Gravity resolution}
\label{sec:vsGravResolution}
Here we test the impact of the gravity field resolution (meaning the maximum degree of the observed Stokes coefficients) on the accuracy of the retrieved interior model. The maximum angular resolution that can be expected from a truncated spherical harmonics series can be as a rule of thumb taken to be that of the half-wavelength of its maximum degree, hence given by $180^{\circ}/l_{\max}$ \citep[e.g][]{jekeliPotentialTheoryStatic2007}. Therefore, a degree-11 gravity would be unable to discern between anomalies at a resolution below about $16^{\circ}\times16^{\circ}$ in latitude and longitude. As for the radial resolution, it is virtually limited by our interior grid discretization, although degree-depth relationship can be found in the literature proving that lower degrees are more sensitive to deeper anomalies \citep{goossensGravityDegreeDepth2023}. 
We therefore expect models obtained with lower degrees to have a larger uncertainty associated with the estimated anomalies shape. Similarly, since the number of measurements decreases when $l_{\max}$ decreases while the number of parameters remains roughly the same, we expect that at each iteration the null space of the locally linearized problem has larger dimensions for lower $l_{\max}$, meaning that a wider range of corrections could provide the same decrease in the objective function. \\

\begin{figure}[!htbp]
      \begin{center}
        \includegraphics[width=\textwidth]{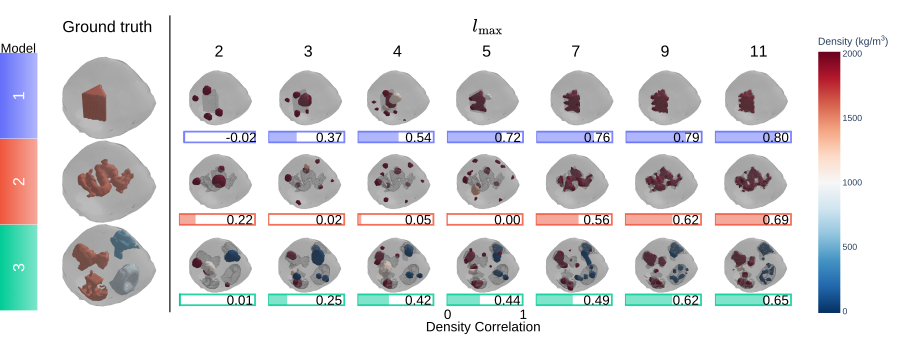}
      \end{center}
      \caption{Density solutions (3D plots) and corresponding Pearson correlations (bar plots) with the true model as a function of the maximum gravity degree of the synthetic noise-free measurement set. The density colorscale is shared between all 3D plots of a given subpanel}
      \label{fig:vsGrav}
\end{figure}

Figure \ref{fig:vsGrav} shows both visually and through the density correlation the behaviour of the GILA single-run solutions with the decrease of the gravity field resolution, for the three reference models. Apart from the measurements resolutions, all settings are maintained from the cases shown in Figure \ref{fig:referenceModels}, meaning we still consider a Bennu-shaped body and measurements unperturbed by noise. The $\chi_P^2$, not shown here, are all close to 1.
It can be seen that for all three models, degree 7 provides a quality of the reconstruction comparable to that obtained from higher degrees. Degree 5 seems to be the threshold for an accurate detection of these models, save for Model 2, which compared to the other two has heterogeneities mostly concentrated near the center of the body.  \\
From these plots, we can empirically define a correlation threshold of 0.2 to indicate a successful reconstruction, with values higher than 0.6 representing an accurate retrieval of the target model. However, as the 3D plots show, even for solutions with correlations lower than 0.2 the estimated density distribution may not be extremely different from the truth, at least in terms of location and sign of the anomaly. 

\subsubsection{Noise}
\label{sec:vsNoise}
\begin{figure}[!htbp]
      \begin{center}
        \includegraphics[width=\textwidth]{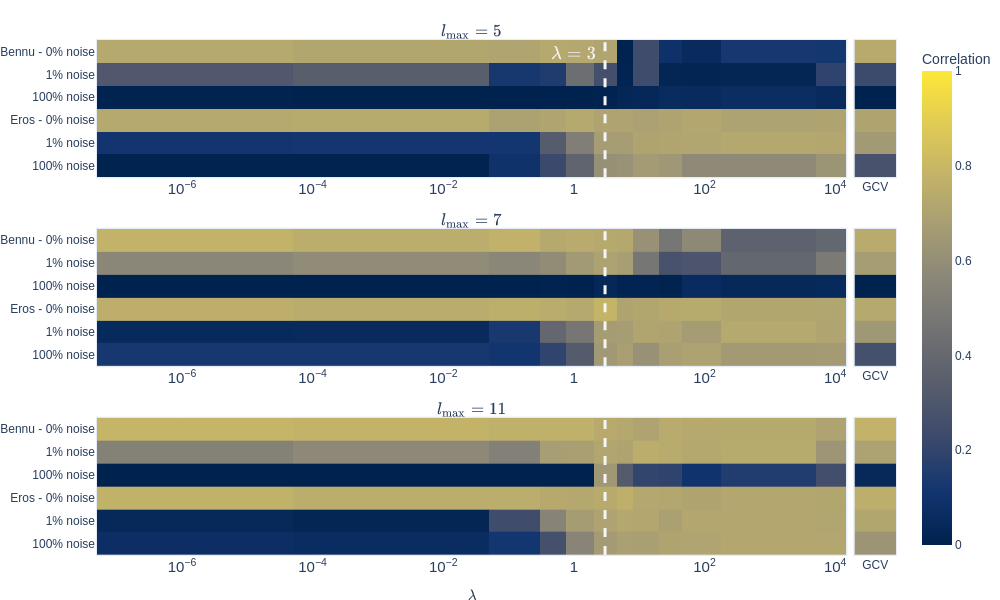}
      \end{center}
      \caption{Correlations of simulation solutions as a function of the regularization settings, for different noise levels, bodies, and gravity resolutions. The first column is the classical Tikhonov regularization for $\lambda$ in the range [$10^{-7},10^{4}$], with the vertical dashed line covering values for $\lambda=3$. The second column is the Tikhonov solution with $\lambda$ selected by generalized cross-validation.}% The third column is the non-regularized solution, only shown for the noise-free case. \slm{Why only noise-free? Like that it looks like non-regularised is the best setting}}
      \label{fig:vsRegParam}
\end{figure}

In this section we examine the effect of data noise on the estimated density distribution. To this end, the measurements are perturbed with zero-mean Gaussian noise, sampled randomly from the profile of Eq. \ref{eq:noiseSpectrum}. 
Small perturbations in the measurements due to noise may result in vastly different least-squares solutions $\bm{\delta t}$, given the inherent instability of the problem. In a singular-value-decomposition (SVD) formulation of the least-squares solution, this effect can be traced to small singular values of the observation matrix $J$ (compared to the highest singular-value) that divide the residuals vector $\bm{y}$, greatly amplifying the errors in the data \citep{bjorckNumericalMethodsMatrix2015}. To mitigate this instability, we add a classical Tikhonov regularization term to the least-squares objective function, minimizing the norm of the correction $\bm{\delta t}$ (second term in Equation \ref{eq:levelSetObjective}). We select an empirical regularization parameter of $\lambda=3$ for both density and level-set corrections. This solution is equivalent to the classical least-squares solution with \emph{a priori} knowledge, where the \emph{a priori} central values of all parameter corrections in $\bm{\delta t}$ are 0 and their \emph{a priori} covariance is a diagonal matrix with diagonal elements equal to $\lambda$, corresponding to \emph{a priori} formal errors of $1/\sqrt{\lambda}$. Figure \ref{fig:vsRegParam} shows the dependence of the solutions on the values of the regularization parameter, justifying our choice of $\lambda$. Each row in the plot corresponds to solutions for the Model 1 ground-truth, two different bodies (Bennu and Eros), and 3 different levels of noise perturbation (0\%, 1\%, and 100\%). For $\lambda<<1$ the solution becomes equivalent to the least-squares solution without regularization. The figure also reports the generalized cross-validation solutions for each case, and their agreement with the $\lambda=3$ solutions, coupled with the heavier computation efforts of the GCV, leads us to prefer the use of a fixed regularization parameter. \\

% \begin{figure}[!htbp]
%       \begin{center}
%         \includegraphics[width=\textwidth]{images/v2/vsRegParam.png}
%       \end{center}
%       \caption{Correlations of simulation solutions as a function of the regularization settings, for different noise levels, bodies, and gravity resolutions. The first column is the classical Tikhonov regularization for $\lambda$ in the range [$10^{-7},10^{4}$]. The second column is the Tikhonov solution with $\lambda$ selected by generalized cross-validation (GCV).}% The third column is the non-regularized solution, only shown for the noise-free case. \slm{Why only noise-free? Like that it looks like non-regularised is the best setting}}
%       \label{fig:vsRegParam}
% \end{figure}

\begin{figure}[!htbp]
      \begin{center}
        \includegraphics[width=\textwidth]{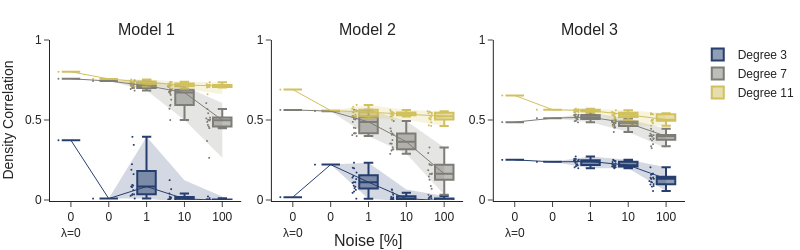}
      \end{center}
      \caption{Density correlations of GILA solutions as a function of the level of data noise contamination, with $\lambda=0$ representing the non-regularized inversions. The points at each noise level represent $\rho_{CORR}$ of solutions for different sets of measurement perturbations sampled within the given noise profile. For each noise level, the box-plots bounds mark the first and third quartile, while their internal line is the median. The length of the whiskers is 1.5 times the interquartile range. For each gravity degree, the lines connect the median values of the density correlations, and the shaded regions cover their range of values.}
      \label{fig:vsNoise}
\end{figure}

The effects of data noise using this regularization approach are presented in Figure \ref{fig:vsNoise}, which shows the evolution of the density correlations for different values of $\alpha$ in Eq. \ref{eq:noiseSpectrum} (corresponding to measurement noise levels of 0\%, 1\%, 10\%, and 100\%), different resolutions of the synthetic gravity measurements ($l_{\max}$ of 3, 7, and 11), and for the 3 target models of Section \ref{sec:targetModels}. Otherwise, all simulation settings are the same as in that section. The set of points labeled as $\lambda=0$ are the solutions using noise-free data and no regularization penalty, which are in general different from the Tikhonov 0\%-noise solutions due to the damping of high-frequency contributions coming from the regularization term. These $\lambda=0$ solutions are only shown for the noise-free case, given their poor handling of noisy data (see Figure \ref{fig:vsRegParam}).
For noise levels higher than 0\%, each point in the plot represents a solution obtained from measurements sampled randomly within the noise profile, meaning each point is obtained from a different random perturbation of the simulated measurements.
The degradation of the retrieved model coming from the contamination of noise in the data is considerable, with most of the Model 1 and 2 solutions in the degree-3 case displaying 0 or negative correlations. However, even with 100\% noise, for degrees 7 and 11 the estimated density distributions preserve a strong correlation (>0.2) with the ground truth. 

\subsubsection{Shape and size of the body}
\label{sec:vsBodyShape}
\begin{figure}[!htbp]
      \begin{center}
        \includegraphics[width=\textwidth]{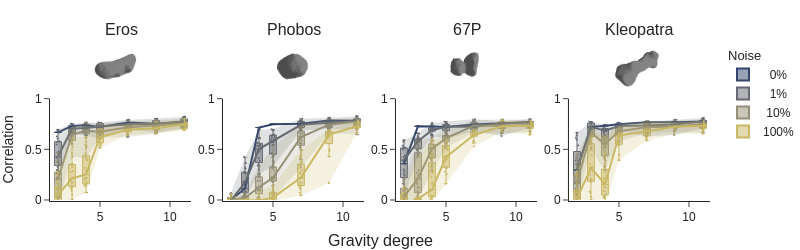}
      \end{center}
      \caption{Density correlation between the Model 1 ground-truth and GILA solutions for simulations over different bodies (one per panel). The points at each noise level represent solutions for different sets of measurement perturbations sampled within the given noise profile. For each gravity degree, the box-plots bounds mark the first and third quartile, while their internal line is the median. The length of the whiskers is 1.5 times the interquartile range. For each noise level, the lines connect the median values of the density correlations, and the shaded regions cover their range of values. The 3D mesh plots show the shape of the body.}
     \label{fig:vsBodyShape}
\end{figure}

So far, we have limited our analysis to a Bennu-shaped body. Now we examine the performance of the model for bodies with different shapes and sizes. The settings are the same as in Section \ref{sec:vsNoise}, with Model 1 as the ground truth and varying gravity resolution and level of noise perturbation, but the input shape model of both the forward and inverse computation is different. 
Figure \ref{fig:vsBodyShape} shows the correlations for the interior densities estimated using shape models of Eros, Kleopatra, Phobos, and comet 67P/Churyumov-Gerasimenko, as a function of the maximum degree of the input gravity measurements and for 4 different noise levels between 0\% and 100\%. The size of these bodies in terms of maximum radius ranges from about \qty{2}{km} for comet 67P to about \qty{60}{km} for Kleopatra, compared to the \qty{290}{m} radius of Bennu. This size difference is not expected to impact the performance of GILA given the normalization of the level-set functions. Indeed, for degrees 5 and higher the correlation of the retrieved model with the ground truth is independent of the body. For degree 3, the reconstruction accuracy is considerably worse for Phobos, as was the case for Bennu (Figure \ref{fig:vsNoise}). We think this could be a consequence of their rounded shapes compared to others of more irregular shape, for which the anomaly would tend to be closer to the surface, where the sensitivity of gravity measurements is higher. 

\subsubsection{Shape error}
\label{sec:vsShapeError}
\begin{figure}[!htbp]
      \begin{center}
        \includegraphics[width=\textwidth]{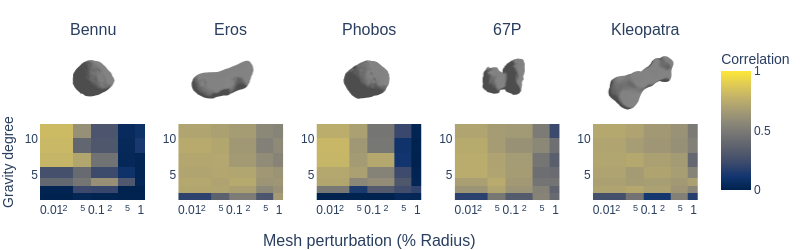}
      \end{center}
      \caption{Influence of shape error on simulated retrievals for Model 1 and noise-free measurements. Each panel displays a heatmap of correlation values between the estimated density distribution and the ground-truth for a specific body shape. The x-axis represents the 1$\sigma$ mesh perturbation applied before the inversion, and the y-axis the maximum gravity degree of the synthetic measurements.}
       \label{fig:vsMeshPerturbation}
\end{figure}

In all previous examples, we assumed a perfect knowledge of the shape, meaning that the same surface mesh used for the data simulation was also used in the inversion. In reality, the shape model of a body has a certain uncertainty associated with it, both coming from the reconstruction errors and from the polyhedral approximation. For Phobos, the 1$\sigma$ error on the shape model can reach \qty{40}{m}, corresponding to about 0.3\% of its radius \citep{willnerPhobosShapeTopography2014a}. For Bennu, the shape error relative to the radius is of similar magnitude, at about 0.25\% \citep{asadValidationStereophotoclinometricShape2021}. Here, we simulate an uncertain knowledge of the shape of the body by perturbing the location of the mesh vertices by a random offset along their normals. The offset for each point is sampled from a normal distribution of given standard deviation. The perturbed model is then used in the inversion, fitting data generated with the original shape model. The results are shown in Figure \ref{fig:vsMeshPerturbation} for a Model 1 ground-truth and noise-free measurements, and shape-offset standard deviations ranging between 0.001 and 1\% of the body radius. As a perturbation on the shape model becomes effectively a perturbation on the residuals, Tikhonov regularization is here necessary to stabilize the inversion.

Overall, GILA solutions show strong correlation with the ground-truth even in the presence of moderate error on the shape. As in Section \ref{sec:vsBodyShape}, the overall behaviour is better for the more irregular bodies, with high correlations even at 1\% mesh perturbation levels and down to degree-3 gravity. For Phobos, as for Bennu, the 0.3\% level appears to be at the edge of detectability for this simulation case.

\subsubsection{Grid discretization}
\label{sec:vsInteriorGrid}
\begin{figure}[!htbp]
      \begin{center}
        \includegraphics[width=0.5\textwidth]{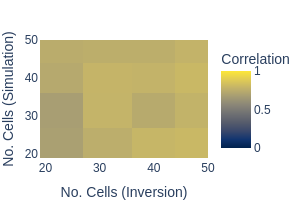}
      \end{center}
      \caption{Heatmap of correlation between reconstructed and ground-truth model. On the x-axis is the resolution of the interior grid used for the inversion, and on the y-axis that of the grid used in simulation, both in terms of number of cells per dimension.}
      \label{fig:vsGridSize}
\end{figure}

Since in the previous cases the interior grid used in simulation was the same as that used in the inversion (same cell size), the target anomaly could in theory be perfectly represented by the inversion model. Of course, in a real case the interior discretization would be another source of error in the reconstruction of the real anomaly. We find however this error to be negligible for GILA, as shown in Figure \ref{fig:vsGridSize}. This heatmap shows the correlation between the inverted model and the ground truth as a function of the resolution of the grid used for the computation of synthetic measurements (in terms of number of grid cells per dimension) and that used in the inversion. The settings are those of Section \ref{sec:targetModels}, with Model 1 as a ground-truth. The correlation is strong for all the cases, with minimum values around 0.6 for coarser models in both simulations and estimation.

\subsubsection{Size, depth, magnitude of the anomaly}
\label{sec:vsAnomalyDepth}
\begin{figure}[!htbp]
      \begin{center}
        \includegraphics[width=0.7\textwidth]{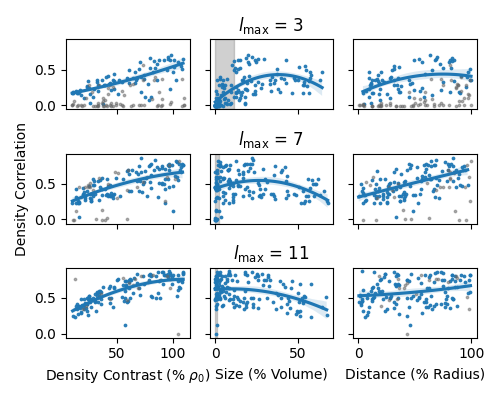}
      \end{center}
      \caption{Scatter plot of density correlations of GILA solutions with their respective ground-truth with a prismatic anomaly of random size, depth, and density contrast. The x-axes represent: the density contrast of the target anomaly relative to the background density $\rho_0$ (first column); the size of the anomaly as a percentage of $\mathcal{V}_\mathcal{B}$ (second column); the distance of the target anomaly from the center of the body (closely related to its depth for a Bennu-shaped body) and relative to the radius of the body (third column). The blue curves represent degree-2 polynomial regressions over the blue points, the shaded areas being their 1$\sigma$ error bounds. The grey points, excluded from the fit, represent solutions where the volume of the target anomaly is lower than $1/{l_{\max}}^2$ times the body's volume (grey boxes in the second column). The rows represent the maximum gravity degree of the synthetic Stokes coefficients, which are contaminated with 100\% Gaussian noise.}
      \label{fig:paper_vsDepth_regPlot}
\end{figure}

In this section we explore the range of detectability of an anomaly in terms of its size, the magnitude of its density contrast, and its depth within the body. To this end, we consider ground-truth distributions based on Model 1, where this time the prismatic anomaly is randomly modified in the location of its center (along the x-axis, with x-coordinate in the range [0, $r_0$]), its density contrast (in the range [$0.1\rho_0$, $1.1\rho_0$]), and its size (scaling factor in the range [$0.06r_0$, $r_0$]), all 3 parameters being sampled from uniform probability distributions. 
The solutions corresponding to each random target model are shown in Figure \ref{fig:paper_vsDepth_regPlot}, for 3 levels of gravity resolution and against the density contrast, size, and distance from the body center of the prismatic anomaly in the true model. The synthetic measurements are here perturbed by 100\% zero-mean Gaussian noise.
The size of the anomaly is expressed in terms of its volume relative to $\mathcal{V}_\mathcal{B}$, and the skewness towards lower values of its distribution (while the sampling of the scaling factor was uniform) is due to the prism not being fully contained inside the body for higher scaling factors.\\
The curves represent order-2 polynomial regressions over the points in blue, grey points being excluded from the fit. These grey points correspond to ground-truth models where the volume of the anomaly relative to $\mathcal{V}_\mathcal{B}$ is lower than $1/{l_{\max}}^2$. This threshold, represented by grey boxes in the figure, is the fraction of the volume of a sphere occupied by a tesseroid of $180^{\circ}/l_{\max}$ in latitude and longitude, which is the half-wavelength spatial resolution of the degree-$l_{\max}$ gravity field (Section \ref{sec:vsGravResolution}). Accordingly, the second column in Figure \ref{fig:paper_vsDepth_regPlot} shows that the density reconstruction becomes worse below this threshold, with most solutions showing near-zero correlations with the ground truth at low gravity degrees. The accuracy of the solution also worsens for large sizes of the target anomalies, possibly because there the bulk density (which we use as an initial value for the background density, $\rho_0$) is further away from the background density of the true model. 
Independently of the size and depth, the accuracy of the retrieved model is higher for higher density contrasts of the target anomaly, although for high resolutions of the gravity successful retrievals are common even for density contrasts of 10\% of the background density.  The dependence of the correlation on the depth of the anomaly, here represented by its distance from the center of the Bennu-shaped body, is lower for high degrees. This is presumably because due to the exponential noise profile, a higher resolution of the gravity leads also to a lower relative uncertainty of the low-degree coefficients, which are more sensitive to deep anomalies. Even at degree 3, GILA seems to be able to retrieve anomalies deep inside the body, as long as their size is above the gravity spatial resolution.

\subsubsection{Initial model}
\label{sec:vsInitialModel}
The non-convexity of the objective function means that the solution is strongly dependent on the starting point of the iterative algorithm, that in our case is the initial density model. In a realistic scenario, the core algorithm of GILA should be run several times starting from a wide range of density models. This should provide a wider view of the space of possible solutions, although a complete exploration of the solution space is hindered by both computational efforts and biases in the estimation process which could make some solutions more likely to appear than others. The aim of this section is to provide an example of how changing the initial model can affect the density solution. 
The target model is inspired by those proposed for Bennu by the OSIRIS-REx team \citep{scheeres_heterogeneous_2020}, which were obtained from the inversion of the optical-tracking $\mathcal{O}_{lm}$ dataset and supported by analytical and geophysical considerations. These models are composed of a spherical core with a mass deficit of 6 to 16\% of the total mass, an equatorial bulge with a density 5 to 12\% lower than the bulk density, and a middle layer with a density 8 to 17\% higher than the measured bulk density of \qty{1190}{kg/m^3}. Consequently, in this simulation campaign we consider as ground-truth a model with a spherical cavity at the center of radius \qty{116}{m} (40\% of the total radius) and an equatorial ring with a negative density jump of \qty{-400}{kg/m^3}, for a background density of \qty{1370}{kg/m^3}, 15\% higher than the bulk density.
We simulate Stokes coefficients up to degree 10, and contaminate them with 100\% noise using as $\beta$ in Eq. \ref{eq:noiseSpectrum} the usual value of 1/3 to be consistent with the previous sections, although Figure \ref{fig:noiseProfilePlot} shows that for the Bennu particle field $\beta$ is closer to 0 (see Figure~\ref{fig:bennuSimUncertainty} and Sec.~\ref{sec:Uncertainties} for discussion relative to $\beta$).  \\
The synthetic measurements are then inverted multiple times, assuming in each run a random initial model, based on variations on the generic initial model employed throughout the previous sections and shown in Figure \ref{fig:referenceModels}. The spheres making up each of the 3 types of anomalies are now randomized in both their number and locations, while their radius is still fixed to 20\% of the body radius. Moreover, the initial background density $\rho_0$ is itself sampled from a uniform distribution between 50\% and 150\% of the bulk density, and the 3 initial density contrasts are sampled from a uniform distribution in the range [-$\rho_0$, $\kappa \rho_0$] ($\kappa=1$ here), allowing for initial models with void regions (large macroporosity).

\begin{figure}[!htbp]
      \begin{center}
        \includegraphics[width=\textwidth]{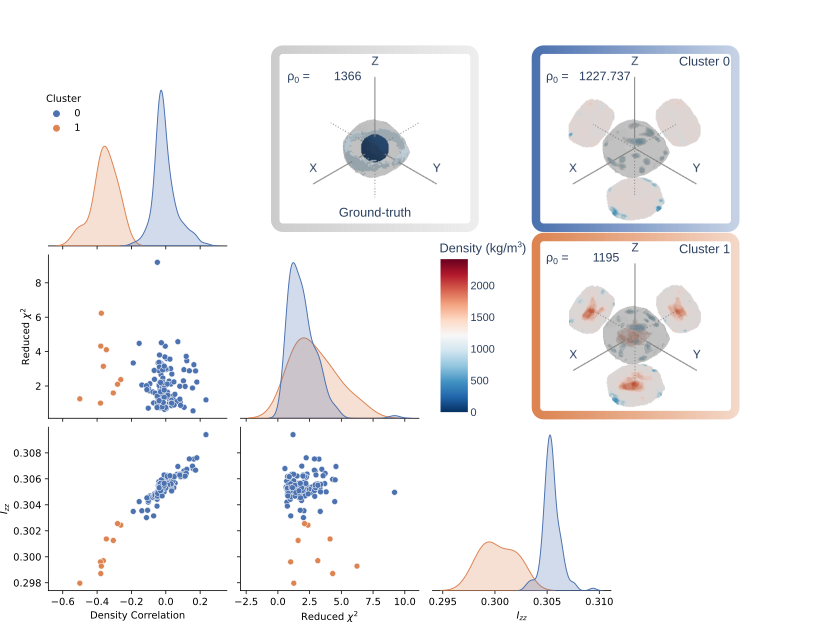}
      \end{center}
      \caption{Density solutions for Bennu's ring-core model from degree-10, 100\% noise data. Bottom left: distributions of the quality metrics and of the normalized moment of inertia $I_{zz}$, with colors indicating the 2 families obtained from clustering of the $I_{zz}$ values. Top right: averages of the density distributions in each cluster (blue and orange boxes), all represented as isosurfaces and slices along the $X=0$, $Y=0$, and $Z=0$ planes, and the ground-truth model (grey box). The value of $\rho_0$ reported above the average solutions is the median of the average density distribution, and the density levels of the 6 isosurfaces are selected from \emph{k-means} clustering of the deviations from this $\rho_0$ throughout the body. The colorbar is relative to the three 3D plots.}
      \label{fig:BennuSims}
\end{figure}

The results are shown in Figure \ref{fig:BennuSims}. For all the distributions the reduced $\chi^2$ value is around 1, indicating a good fit with respect to the data and model errors. As the distribution of the density correlations appears to be multimodal, we separate the full set of solutions into 2 cluster, based on a \emph{k}-means clustering\footnote{\url{https://scikit-learn.org/stable/modules/generated/sklearn.cluster.KMeans.html}} of the values of the (normalized) principal moment of inertia along $z$ over all the density models. The moment of inertia $I_{zz}$ is a global property of the density distribution which is independent of the gravity data used to produce the interior model, since only the difference between any 2 principal moments of inertia is constrained by the gravity \citep{le_maistre_signature_2019}. For this reason, it can be used to differentiate between 2 models which both fit the gravity measurements at the same level. More generally, the multiple solutions of the gravity inversion can be grouped based on the $l^{th}$-order moments of the density function, defined as \citep{jekeliPotentialTheoryStatic2007}:
\begin{equation}
\label{eq:densityMoments}
\mu_{abc} = \int_{\mathcal{V}_\mathcal{B}}x^{a}y^{b}z^{c}\rho(\bm{x})dV
\end{equation}
with $a,b,$ and $c$ integer indices such that $a + b+ c=l$. The moments of order $l$ are related to the Stokes coefficients of degree $l$, but not all the $ (l_{\max}+1)(l_{\max}+2)(l_{\max}+3)/6$ moments up to order $l_{\max}$ are unambiguously determined by the ${(l_{\max}+1)}^2$ Stokes coefficients up to degree $l_{\max}$ \citep{tricarico_global_2013} and thus constrained by the gravity measurements.\\

Figure \ref{fig:BennuSims} shows a representative model for each cluster, obtained as the average of the density values at each cell over all solutions. The best set of models, shown in blue, includes about 93\% of the 146 solutions, and has a mean correlation of about 0.1. While the value itself is low (lower than the identified 0.2 threshold), it can be seen from the average model that in general the solutions in this group do present a lighter outer region at the equator and negative anomalies near the center, where the void core is supposed to be. The remaining solutions, shown in orange, have a low moment of inertia and negative correlation, both justified by the presence of positive anomalies deep within the body which appear in the average solution for the cluster. Overall, the concentration of most solutions around a single cluster may suggest an incomplete exploration of the solution space, to be ascribed either to a range of initial models not diverse enough, or to considerable biases in the inversion process. 

\subsection{Uncertainty estimation}
\label{sec:Uncertainties}
We characterize here possible approaches to the estimation of errors associated with a retrieved density distribution. The non-convexity of the problem and the under-constrained approach we chose mean the formal errors provided by the least squares method are not meaningful quantities. A linear estimation of the density fixing the shapes at convergence could provide error bars for the density contrasts of each anomaly, but these will be specific for the estimated shape and not representative of the high correlation between anomaly shape and density jump, so probably still too optimistic. We therefore appraise the uncertainty associated with each solution by analyzing the spread of different runs converging to the same interior family (cluster), similarly to \citet{dinsmoreConstrainingInteriorsAsteroids2023}. In simulation, where the ground-truth is known, these uncertainties can be compared to the true errors to assess their reliability. A set of representative solutions can be generated by starting from different initial models, as in Section \ref{sec:vsInitialModel}, or by randomly perturbing the observed measurements within their associated noise, or by a combination of the two approaches. \\

\begin{figure}[!htbp]
      \begin{center}
        \includegraphics[width=\textwidth]{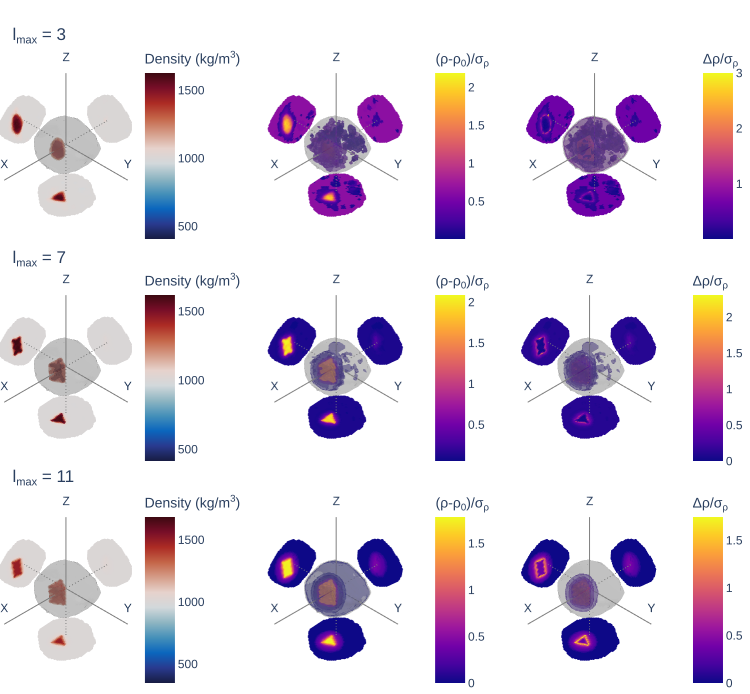}
      \end{center}
      \caption{Average models and associated density uncertainties for multiple synthetic retrievals based on the Model 1 ground truth for Bennu. Only results relative to the largest cluster are shown here. The rows indicate the $l_{\max}$ of the simulated noise-free $\mathcal{O}_{lm}$. The first column shows the average solutions. The second column shows the significance of the detected density anomaly, as the ratio of the estimated deviation from the background density and the computed density standard deviation. The third column shows the reliability of the estimated uncertainties, as the ratio of the true errors (average density model minus the ground-truth density) and $\sigma_{\rho}$. Values larger than 3 in this column are set to 3. The 3D mesh plots all represent isovalues of the quantity surrounded by slices along the $X=0$, $Y=0$, and $Z=0$ planes} 
      \label{fig:densityUncertainty}
\end{figure}

Figure \ref{fig:densityUncertainty} shows the average solution and corresponding errors for simulation runs with the Model 1 ground-truth, assuming no noise and 3 levels of gravity resolution. Each plot is obtained from statistics over about 200 converged solutions, each starting from a different random initial model as described in Section \ref{sec:vsInitialModel}. As shown in that section, we additionally cluster the full set of solutions based on the values of $I_{zz}$, using 2 cluster centers. Most of the solutions (95\%) are contained in the same cluster, which has high correlation with the ground-truth model. We then compute the uncertainty ($\sigma_{\rho}$) as the standard deviation of the distribution of density values of each cell over the set of solutions in a given cluster, while the average solution is given by the mean of these distributions. We plot the ratio between the density contrast of the average solution (difference between the cell density and the median density over the body) and the computed uncertainties, thus representing the statistical significance of any detected heterogeneity. We see that inside the target anomaly the density jump is statistically significant, with the errors generally about 60\% of the deviation. The confidence of the average solutions is lower at the edges of the target anomaly, where the density contrasts are generally consistent with 0. This is explained by the average solution coming from models where the retrieved anomaly is smaller and of larger density jump compared to the truth, or larger and with a lower density jump. Near the center of the true anomaly, all these solution will have a positive density jump, while around the border the effect of the denser and the lighter anomalies will cancel out. The last column in Figure \ref{fig:densityUncertainty} shows the ratio of the absolute values of the true errors ($\Delta \rho = \rho^{(est)}-\rho^{(sim)}$) and the density standard deviation. We see again that near the center of the anomaly the uncertainties are larger than the true errors, while at the borders they tend to slightly underestimate the true errors in the solution. \\

\begin{figure}[!htbp]
      \begin{center}
        \includegraphics[width=\textwidth]{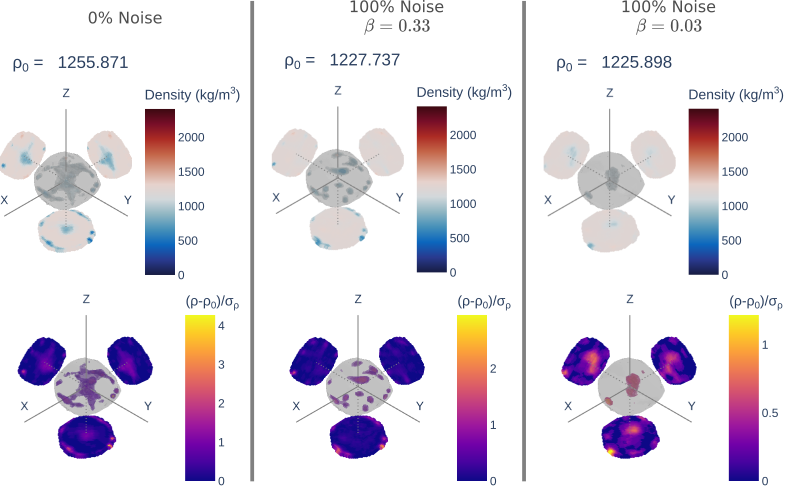}
      \end{center}
      \caption{Average models and associated density uncertainties for multiple synthetic retrievals based on Bennu's ring-core model of Section \ref{sec:vsInitialModel}. The rows indicate the average model and its significance, computed as the ratio of its deviation from its median ($\rho_0$) and the standard deviation of the computed density. The first column was obtained from noise-free measurements, while for the second and third columns the synthetic measurements were perturbed following a 100\% noise profile with $\beta = 1/3$ and $\beta = 0.03$, respectively. The 3D mesh plots all represent isovalues of the quantity surrounded by slices along the $X=0$, $Y=0$, and $Z=0$ planes. In the second row, the isovalues are only shown for values larger than 0.5.} 
      \label{fig:bennuSimUncertainty}
\end{figure}

Similar statistics can be computed for the set of solutions relative to the scenario of Section \ref{sec:vsInitialModel} with the ring-core model. Figure \ref{fig:bennuSimUncertainty} shows the average models and uncertainties obtained from the ring-core ground-truth, with $l_{max}=10$ and for 3 different measurement noise settings: noise-free measurements, 100\% noise with $\beta = 1/3$ (same as Figure \ref{fig:BennuSims}), and 100\% noise with $\beta = 0.03$ (approximating the OSIRIS-REx particle field profile). Each solution within a set is obtained by starting from a random initial model, but the values of the observables are the same in each set (same measurement perturbation), since the aim of Section \ref{sec:vsInitialModel} was to gauge the influence of the initial density distribution. For each of these cases, the statistics are relative to the cluster with highest mean correlation. With perfect measurements GILA is able to retrieve the under-dense ring and central core with statistical significance. The ring is also present in the more realistic $\beta = 1/3$ case and with a density contrast on average larger than the corresponding uncertainty, but the central under-dense region corresponding to the core in the true model is less well defined. The $\beta = 0.03$ average model presents the inner lighter regions and a small portion of the equatorial ring, but both anomalies are almost consistent with 0, meaning that a fully homogeneous distribution could not be ruled out based on these results.

\subsection{Realistic cases}
\label{sec:realisticCases}
The two hypotheses of piece-wise constant distribution and low number of distinct anomalies may appear to severely limit the range of applicability of GILA in real estimation cases. However, we show in the following simulations that the method could converge to a reasonable approximation of the true model even when the ground-truth does not respect one of these two assumptions. Specifically, we treat here the case where the true density distribution is a smooth function, and the case of a rubble-pile body, where the larger rocks and voids could be considered anomalies dispersed in a background of finer material.
\subsubsection{Continuous target distribution}
\label{sec:PhobosPC}
\begin{figure}
     \centering
     \begin{subfigure}[t]{0.49\textwidth}
         \centering
         \includegraphics[width=\textwidth]{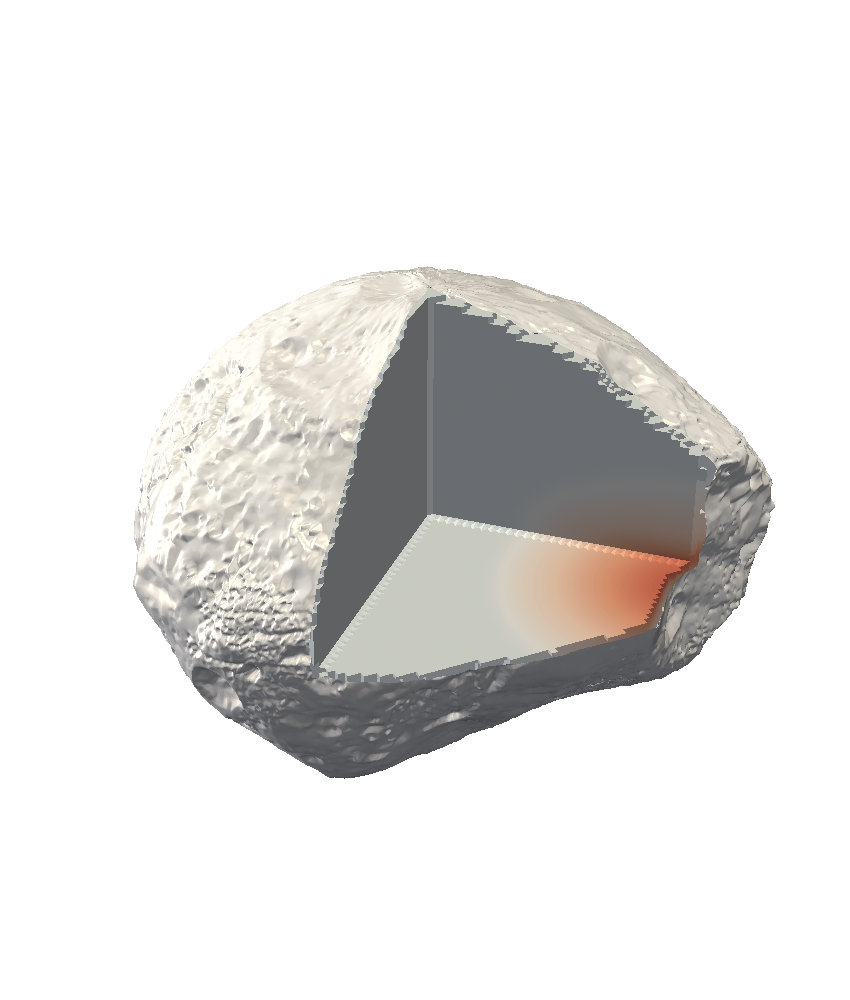}
         \caption{Smooth density distribution for the Phobos porous compressed model}
         \label{fig:phobosPCFw}
     \end{subfigure}
     \hfill
     \begin{subfigure}[t]{0.49\textwidth}
         \centering
         \includegraphics[width=\textwidth]{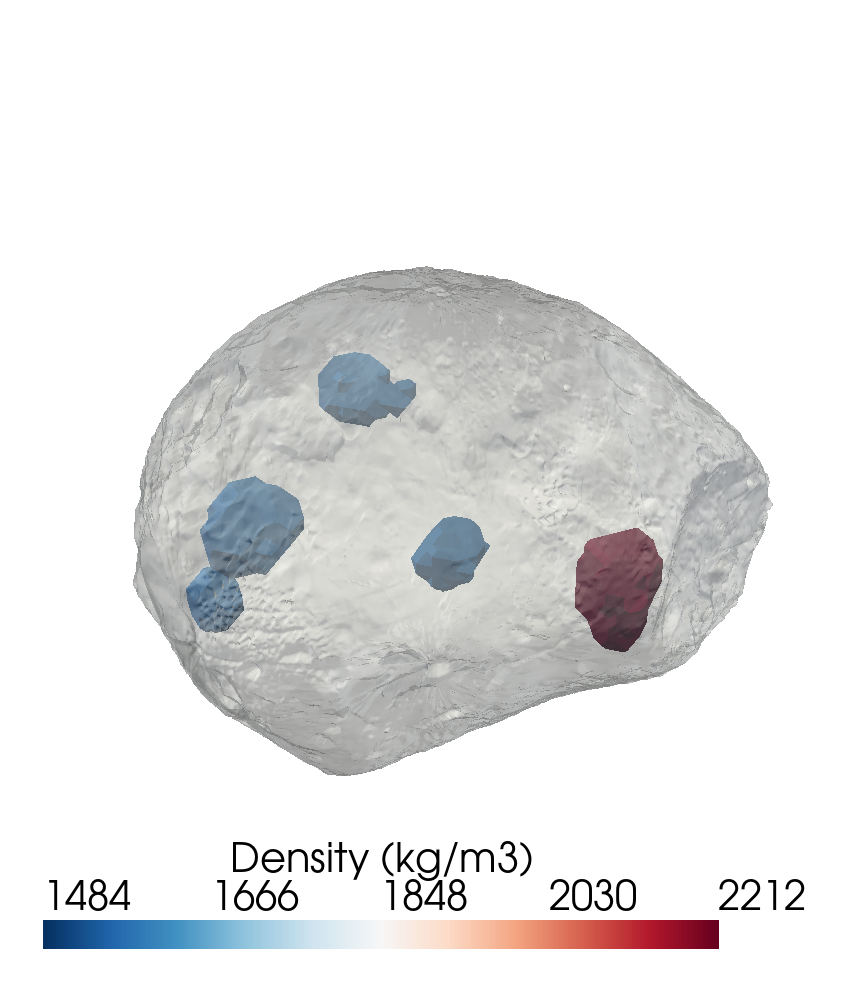}
         \caption{Estimated distribution from a degree-5, 100\%-noise gravity}
         \label{fig:phobosPCInv}
     \end{subfigure}
     \vskip\baselineskip
     \begin{subfigure}[t]{0.49\textwidth}
         \centering
         \includegraphics[width=\textwidth]{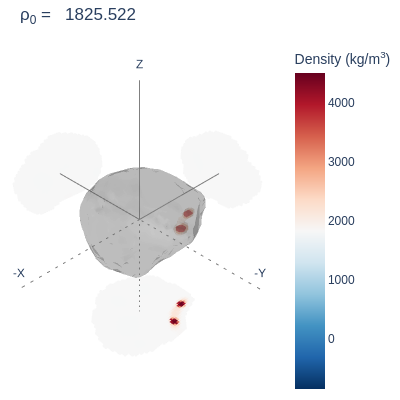}
         \caption{Average of 200 density solutions from the same degree-5, 100\%-noise gravity and random initial model.}
         \label{fig:phobosPCAvg}
     \end{subfigure}
     \hfill
     \begin{subfigure}[t]{0.49\textwidth}
         \centering
         \includegraphics[width=\textwidth]{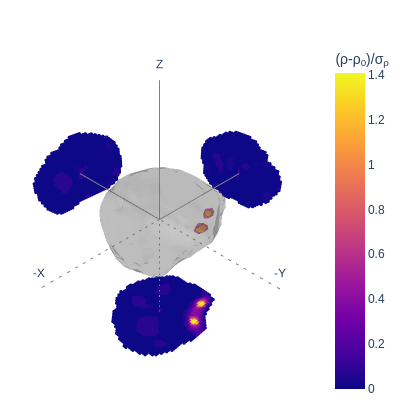}
         \caption{Density contrast of the average solution with respect to its median value ($\rho_0$), divided by the corresponding standard deviation of the 200 solutions at each cell.}
         \label{fig:phobosPCSigma}
     \end{subfigure}
    \caption{Forward model (a), single-run solution (b), average solution (c) and its significance (d), for the Phobos porous compressed simulation case. The 3D mesh plots in the top row share the same colorbar. The 3D mesh plots in the bottom row all represent isovalues of the quantity surrounded by slices along the $X=0$, $Y=0$, and $Z=0$ planes. For panel (d), the isovalues are only shown for values larger than 0.5.}
    \label{fig:mixedEst}
    
\end{figure}

A smooth variation of the density within a small body could for example be formed by compression of a porous material following an impact. This is the characteristic of one of the interior families proposed by \citet{le_maistre_signature_2019} for Phobos, the "Porous Compressed" (PC) model. Based on the values of Table 2 in that paper, we construct a member of the PC family by placing right under the Stickney crater an ellipsoid of semi-major axes (\num{7.5}, \num{7.5}, \num{3.5}) \unit{km} in body-fixed Cartesian coordinates. The density contrast inside the anomaly is then obtained by scaling the signed-distance function from its border so that the maximum value is \qty{200}{kg/m^3}. Therefore, the density contrast decreases smoothly away from this peak down to \qty{0}{kg/m^3} at the border of the ellipse, where the total density is the same as that outside the anomaly, namely \qty{1848}{kg/m^3}. In practice, inside the anomaly the density is not exactly smooth due to the interior grid discretization, for which however the resolution is relatively high, at 100 cells per dimension.
Figure \ref{fig:phobosPCFw} shows the resulting ground-truth model. Hence, from this distribution we produce synthetic measurements of $l_{\max}=5$ and 100\% noise level. The corresponding GILA distribution estimated over a grid of 50 cells per dimension is shown in Figure \ref{fig:phobosPCInv}, where the initial model for this single run of the algorithm is the same as in Section \ref{sec:targetModels}. There is indeed a positive density anomaly found inside the original ellipsoid, although off-centered, and smaller with about twice the density jump. On the other hand, there are also multiple negative anomalies on the opposite side of the body, with a density contrast of roughly \qty{-200}{kg/m^3}. To gauge the statistical significance of all these anomalies, we perform multiple inversions from random models as in Section \ref{sec:Uncertainties}. The average of the 200 solutions is shown in Figure \ref{fig:phobosPCAvg}, where the positive anomaly below Stickney (+X, -Y direction) is clearly visible, although its density contrast is up to 10 times the maximum density contrast of the anomaly in the true model. Figure \ref{fig:phobosPCSigma} shows the ratio of the density contrasts in the average model and their spread, computed as the standard deviation $\sigma_{\rho}$ of the density at each cell across the solutions, as in Section \ref{sec:Uncertainties}. The density jump of the detected anomaly is about 1.4 $\sigma_{\rho}$ away from 0, making it statistically significant.

\subsubsection{Rubble-pile target}
\label{sec:rubblePile}
\begin{figure}
     \centering
     \begin{subfigure}[t]{0.32\textwidth}
         \centering
         \includegraphics[width=\textwidth]{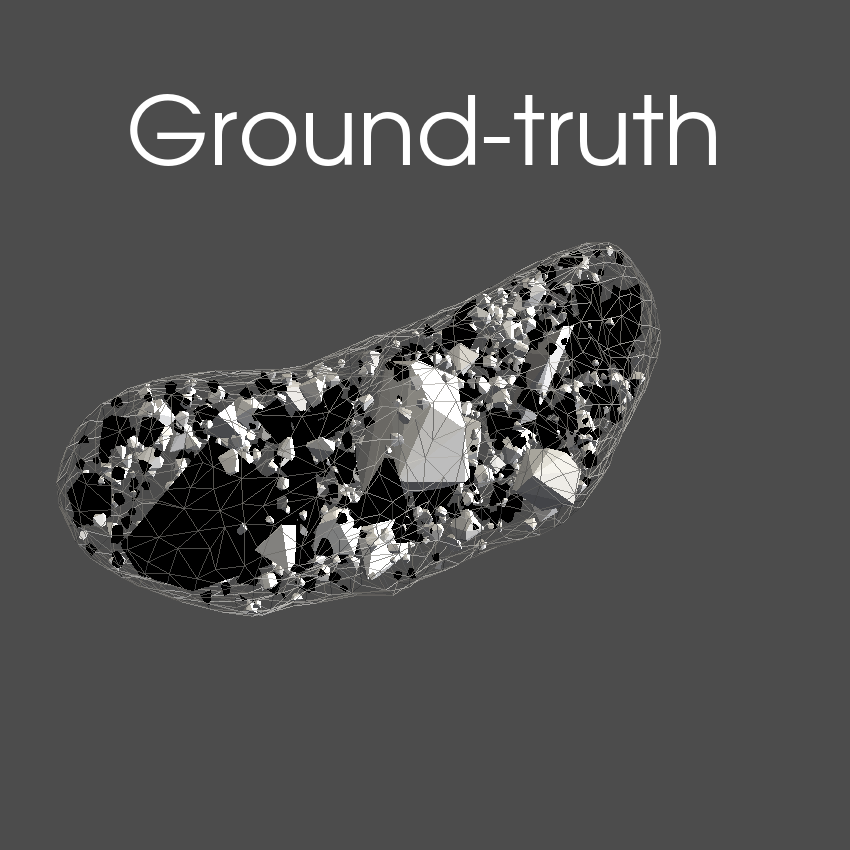}
         \caption{Random rubble-pile distribution for an Eros-shaped body.}
         \label{fig:rubbleFw}
     \end{subfigure}
     \hfill
     \begin{subfigure}[t]{0.32\textwidth}
         \centering
         \includegraphics[width=\textwidth]{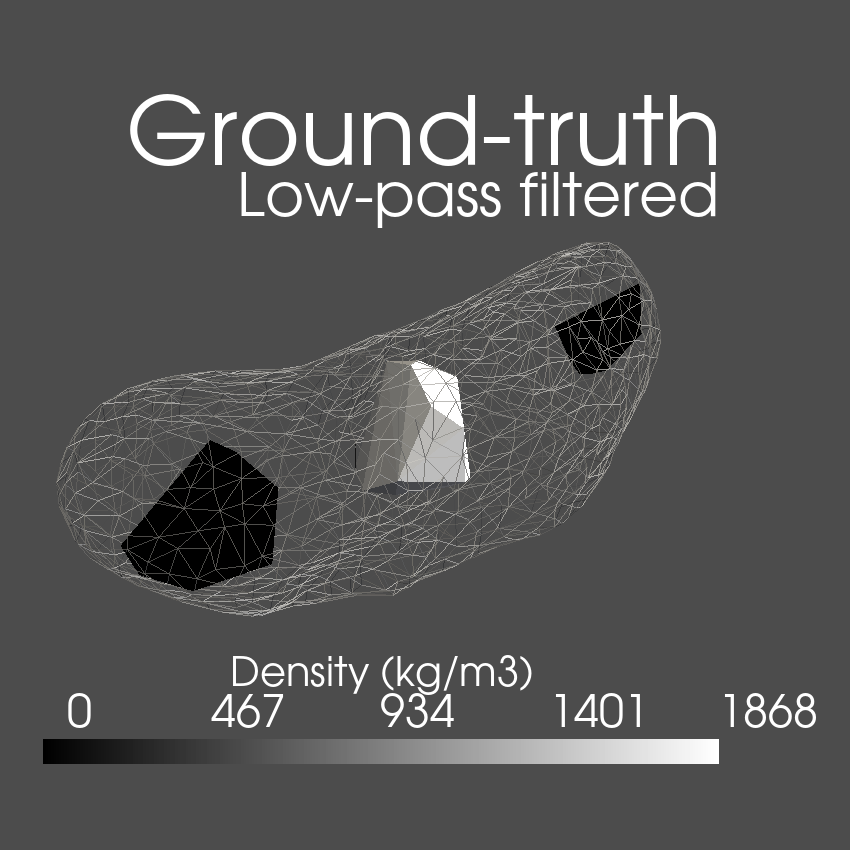}
         \caption{Ground-truth model when only elements with relative volume larger than $1/l_{\max}^2$ are plotted.}
         \label{fig:rubbleFwLPF}
     \end{subfigure}
     \hfill
     \begin{subfigure}[t]{0.32\textwidth}
         \centering
         \includegraphics[width=\textwidth]{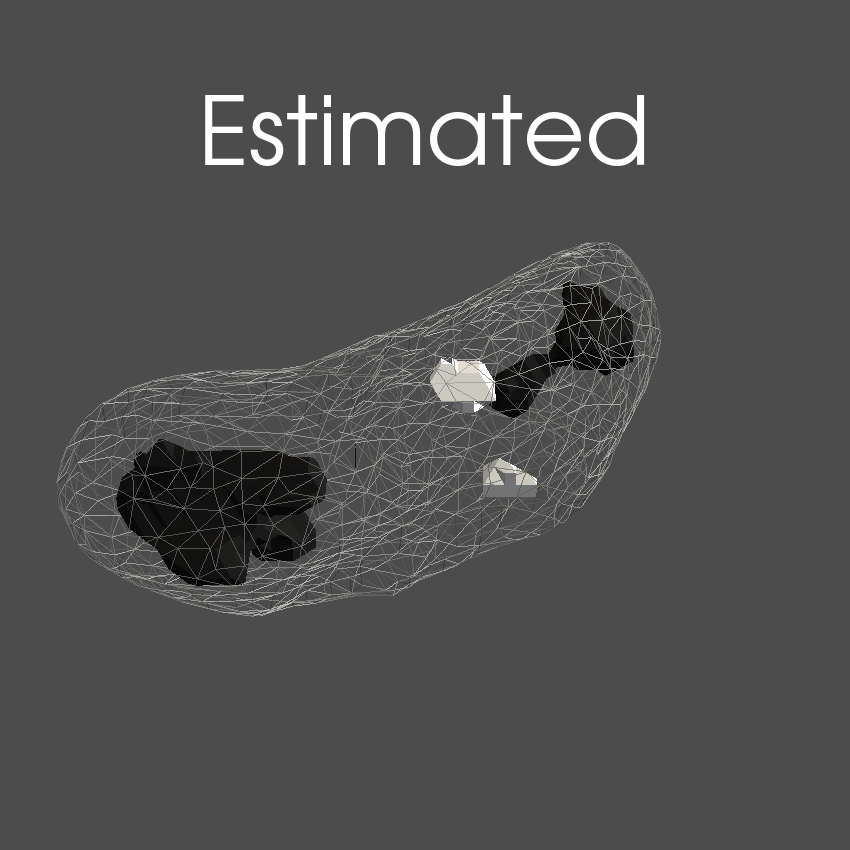}
         \caption{Estimated distribution from a degree-10, 100\%-noise gravity}
         \label{fig:rubbleInv}
     \end{subfigure}
     \vskip\baselineskip
     \begin{subfigure}[t]{0.49\textwidth}
         \centering
         \includegraphics[width=\textwidth]{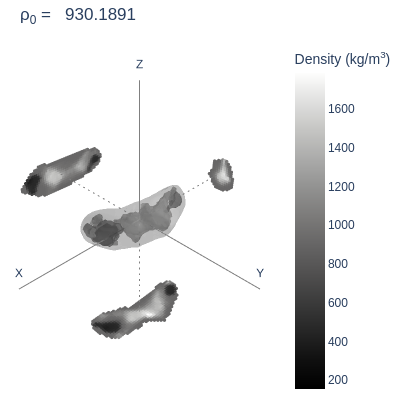}
         \caption{Average of 200 density solutions from the same degree-5, 100\%-noise gravity and random initial model.}
         \label{fig:rubbleAvg}
     \end{subfigure}
          \hfill
     \begin{subfigure}[t]{0.49\textwidth}
         \centering
         \includegraphics[width=\textwidth]{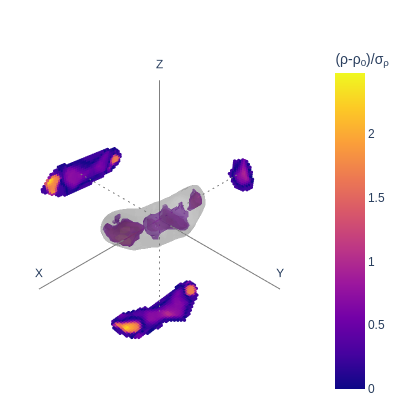}
         \caption{Density contrast of the average solution with respect to its median value ($\rho_0$), divided by the corresponding standard deviation of the 200 solutions at each cell.}
         \label{fig:rubbleSigma}
     \end{subfigure}
    \caption{Forward model (a) and its low-pass-filtered visualization (b), single-run solution (c), average solution (d) and its significance (e), for the Eros-shaped rubble pile simulation case. The 3D mesh plots in the top row share the same colorbar. The 3D mesh plots in the bottom row all represent isovalues of the quantity surrounded by slices along the $X=0$, $Y=0$, and $Z=0$ planes. For panel (e), the isovalues are only shown for values larger than 0.5.}
    \label{fig:test}
    
\end{figure}

Formation and evolution models suggest that most of the asteroids with sizes between \qty{200}{m} and \qty{10}{km} are rubble-piles \citep{walshRubblePileAsteroids2018}. We therefore test here the ability of the GILA to retrieve an approximate density distribution when the target body is a rubble-pile. We still assume that the body is composed of a uniform background density, which could model the finer particles and corresponding voids while allowing the shape to be perfectly filled. In this matrix of constant density are dispersed random blocks of larger size ($>1\%$ of the body radius). We model the polydisperse rubble pile interior with a power-law size-frequency distribution (SFD) for the rubble and for the voids, as in \citet{tricaricoInternalRubbleProperties2021}. We select an SFD index of -2 for both the rubble and the voids distributions, so as to generate models with a larger fraction of bigger boulders within the body than what is generally observed on the surface of rubble-piles, where this index is closer to -3 \citep{tricaricoInternalRubbleProperties2021}. Both rocks and voids are modelled as polyhedra, with a random number of vertices between \num{8} and \num{20} whose positions are sampled from the vertices of a spherical mesh of radius given by the SFD. Figure \ref{fig:rubbleFw} shows a randomly generated rubble-pile target of Eros with both rubble, of density \qty{1500}{kg/m^3}, and voids, dispersed in a matrix of density $\rho_0 =\qty{1000}{kg/m^3}$. This ground-truth model is used to generate a set of $\mathcal{O}_{lm}$ coefficients with $l_{\max}=10$. There is no interior discretization grid in this forward computation, instead the total heterogeneous $\mathcal{O}_{lm}$ are computed by summing the $\mathcal{O}_{lm}$ of each polyhedral element, weighted by its excess mass, to the $\mathcal{O}_{lm}$ of the body mesh weighted by the mass of a homogeneous body with density $\rho_0$. \\

According to the results in Figure \ref{fig:paper_vsDepth_regPlot}, the estimation may be inaccurate for elements with volume smaller than $1/l_{\max}^2$ times the body volume, based on the half-wavelength resolution of the gravity coefficients. Figure \ref{fig:rubbleFwLPF} shows only the components of the same ground-truth model with relative size larger than $1/l_{\max}^2$, which can be thought of as a low-pass filter acting on the density distribution. Indeed, only these larger elements are detected by GILA, as can be seen in Figure \ref{fig:rubbleInv}, presenting the density distribution retrieved from the degree-10 gravity perturbed by 100\% noise. As in Section \ref{sec:PhobosPC}, the statistical significance of the estimated anomalies is tested by exploring multiple solutions obtained from the same dataset but starting from different initial models. Clustering was not performed here, the distribution of the $I_{zz}$ being close to unimodal. The averaged solution is plotted in Figure \ref{fig:rubbleAvg}, where the two negative anomalies and an overdense region in the central part are still discernible. However, Figure \ref{fig:rubbleSigma} shows that while the negative anomalies have associated errors which are less than half their density contrast, a large part of the positive anomaly at the center has error bars covering their full density contrast, making it consistent with an absence of heterogeneity. Still, GILA appears to be able to retrieve approximations to complex density models, within the resolution of the gravity field.

%\section{Application to Bennu}}
%\label{sec:realData}
%\subsection{\slm{Simulation study}}
%\slm{Show here the case with the torus, using the central value of the distributions of core, torus and mantle of Scheeres et al. - It should work quite well.}\\

\section{Preliminary analysis of Bennu real data}
\label{sec:realData}
We test here the behavior of the method on real measured Stokes coefficient. Specifically, we use the set of Bennu Stokes coefficient estimated by the OSIRIS-REx team from optical tracking of ejecta particles in orbit around the asteroid \citep{chesleyTrajectoryEstimationParticles2020}. The particle field solution includes coefficients up to degree 10, which are also consistent with the lower-resolution ($l_{\max}=\num{4}$) spacecraft-radio-tracking estimates (see Figure \ref{fig:noiseProfilePlot}). As discussed in \citet{scheeres_heterogeneous_2020}, only coefficients up to degree-4 are sensitive to heterogeneities within the body. This means that the difference between the central values of these low-degree coefficients and a constant-density field simulated from the available shape model are still statistically significant given the estimated uncertainties. However, we expect GILA to handle the non-significance of the higher degrees given the inclusion of weights in the least squares, which is why the full set of coefficients is used here in the inversion. On the other hand, although the published particle field has an associated covariance matrix $\Omega$, we here set all correlations to 0 and only use the diagonal part, as GILA was validated on this simpler scenario of independent measurements. Solutions with the inclusion of the full covariance matrix by Cholesky factorization of $\Omega$ (Section \ref{sec:leastSquaresTheory}) show unstable convergence history, possibly expressing the need for more specific regularization which will be the object of future work. The effect of this approximation on our solution is hard to estimate, but however large it may be it leads to an overestimation of the information content of the data. \\

\begin{figure}[!htbp]
      \begin{center}
        \includegraphics[width=\textwidth]{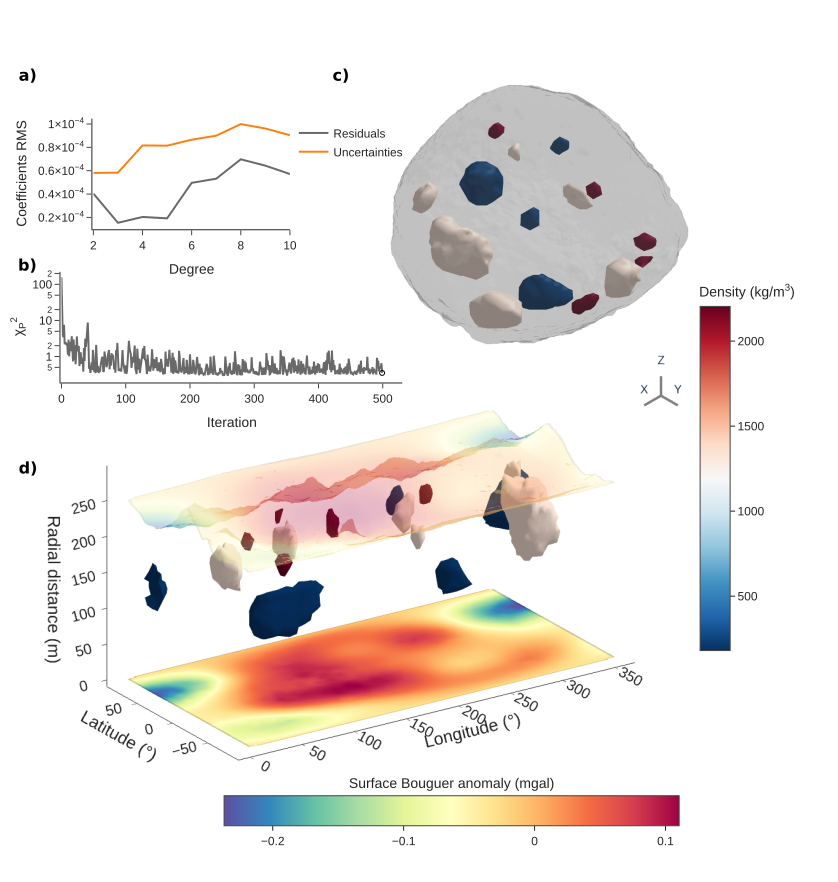}
      \end{center}
      \caption{Density estimation from the OSIRIS-REx Bennu particle field coefficients. RMS of the residuals and the measurement uncertainties for each degree (a). Evolution of the $\chi_P^2$ metric over the 500 iterations (b). Density model at convergence (c). Estimated density distribution in spherical coordinates, with the associated surface Bouguer anomalies both mapped onto the surface of the body (see Figure 2 of \citet{scheeres_heterogeneous_2020}) and also projected on the $r=0$ plane (d). The density colorbar is shared between panels c and d.}
      \label{fig:BennuReal}
\end{figure}

Figure \ref{fig:BennuReal} shows the density distribution estimated from the Bennu particle field, using the SPC v42 shape model \citep{asadValidationStereophotoclinometricShape2021}, and starting from the generic initial distribution shown in Figure \ref{fig:referenceModels}. It mainly features 3 negative anomalies, 1 at the South pole and 2 at mid-latitudes around $0^{\circ}$ and $250^{\circ}$ longitude in the Northern hemisphere, and a concentration of positive anomalies around $120^{\circ}$ longitude in the Southern hemisphere. The heatmap in the bottom panel represents the Bouguer anomalies (difference between the measured and constant-density gravity accelerations) over the surface of the body, as shown in Figure 2 of \citet{scheeres_heterogeneous_2020}. As discussed in \citet{scheeres_heterogeneous_2020}, given that only the heterogeneous component of the Stokes coefficients is used for this plot, the error due to the evaluation of the spherical harmonic expansion at the surface (thus within the Brillouin sphere) is estimated to be below 1\% everywhere. Both the measured and the constant-density sets of coefficients are truncated at degree 4, since above that degree their difference is smaller than the uncertainties of the measured coefficients. The Bouguer anomalies thus computed show a strong correlation with the estimated density anomalies. Figure \ref{fig:BennuReal} also shows the convergence history, with a final $\chi_P^2$ of 0.34, and the RMS of the residuals for each degree, all below the RMS of the measurements uncertainties. It is worth noting that for this dataset the GILA model resolution, as defined in Section \ref{sec:chiSquared} (see Figure~\ref{fig:modelError}), is always below the measurement noise (see Figure~\ref{fig:noiseProfilePlot}), meaning that the measurements uncertainties shown here and the corresponding weights appearing in the computation of $\chi_P^2$ are strictly the uncertainties of the particle field. 

\begin{figure}[!htbp]
      \begin{center}
        \includegraphics[width=\textwidth]{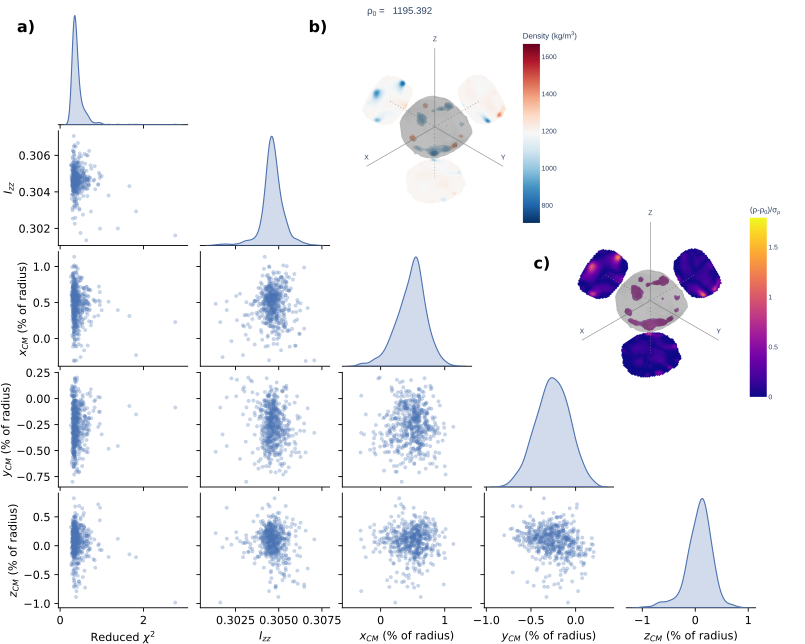}
      \end{center}
      \caption{Density solutions from the Bennu particle field. distributions of reduced $\chi^2$, $I_{zz}$, and location of center of mass (a). Average density distribution (b) and its significance, computed as the ratio of its deviation from its median ($\rho_0$) and the standard deviation of the computed density (c). The 3D mesh plots all represent isovalues of the quantity surrounded by slices along the $X=0$, $Y=0$, and $Z=0$ planes. For panel c, the isovalues are only shown for values larger than 0.5.}
      \label{fig:BennuReal_pairplot}
\end{figure}

The inversion was repeated 500 times using multiple random initial models and perturbing the measurements within their noise, in order to explore the solution space and check for overfitting. Statistics for the resulting models are shown in Figure \ref{fig:BennuReal_pairplot}, including the distribution of their center-of-mass, which was unconstrained here since the degree-1 coefficients were not estimated in the generation of the particle field, but set to 0. As the distribution of $I_{zz}$ and center-of-mass coordinates are close to unimodal, no clustering was performed here. The solutions show high variance, but present the same significant features as the model in Figure \ref{fig:BennuReal}, namely the 2 negative anomalies in the Northern hemisphere and that at the South Pole, and the positive anomalies in the Southern hemisphere.

\begin{figure}[!htbp]
      \begin{center}
        \includegraphics[width=\textwidth]{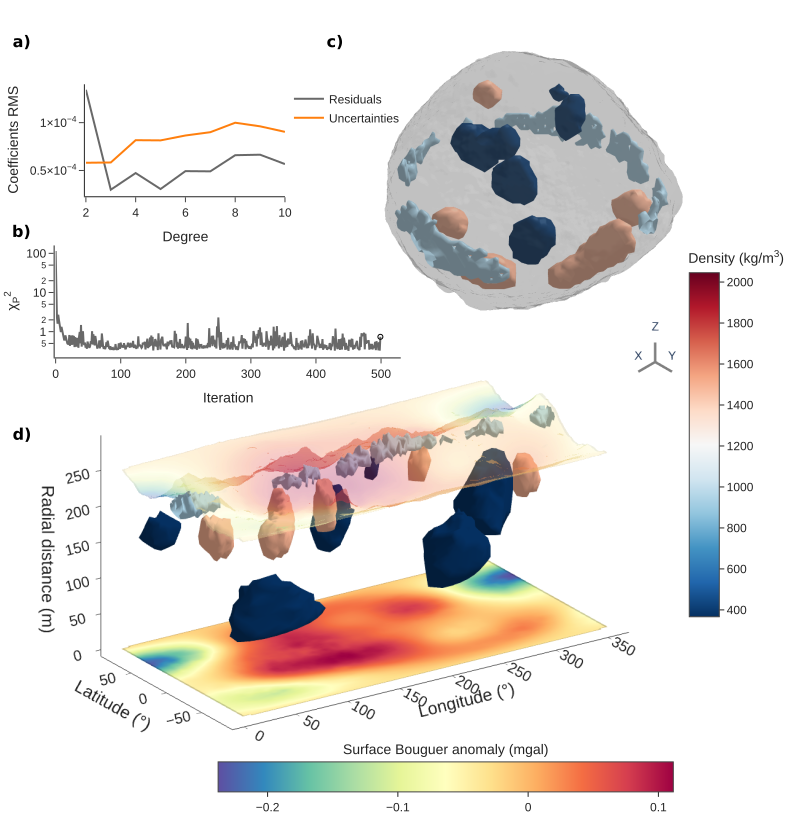}
      \end{center}
      \caption{Density estimation from the OSIRIS-REx Bennu particle field coefficients, with constraints promoting the presence of an underdense ring at the equator. RMS of the residuals and the measurement uncertainties for each degree (a). Evolution of the $\chi_P^2$ metric over the 500 iterations (b). Density model at convergence in Cartesian coordinates (c), and in spherical coordinates along with the surface Bouguer anomalies (d).}
      \label{fig:BennuReal_wConstraints}
\end{figure}

While agreeing with the gravity measurements (within their formal errors and neglecting correlations) and correlating well with the heterogeneous component of the potential, these density distributions are different from the models proposed by \citep{scheeres_heterogeneous_2020} and described in Section \ref{sec:vsInitialModel}. Their ring-core models fit the data equally well, yet GILA was never able to retrieve such a configuration. In fact, all solutions in Figure \ref{fig:BennuReal_pairplot} are similar and essentially belonging to the same family of interior, which could again indicate an incomplete exploration of the solution space. After all, the possibility of retrieving a ring-core model was already tested in simulations and proven to be challenging with settings close to that of the real OSIRIS-REx data ($\beta=0.03$ in Figure \ref{fig:bennuSimUncertainty}). On the other hand, a ring-core model for Bennu is strongly supported by geophysical considerations. The ring corresponds to the intersection of the Roche lobe with the surface of Bennu, where loose material is expected to accumulate \citep{scheeres_geophysical_2016}. The underdense core could as well be justified by failure modes of rubble-pile bodies undergoing spin-up \citep{scheeres_geophysical_2016, zhangInferringInteriorsStructural2022}. The density distributions retrieved by GILA were instead free of any physical assumptions. However, the method supports the addition of quadratic constraints, as explained in Section \ref{sec:leastSquaresTheory}. We therefore add such constraints to favor the presence of an equatorial ring of low density, which is likely for a top-shaped asteroid, by forcing the level-set function of one of the input anomalies to be close to that of the toroidal anomaly used in the ground-truth model of Figure \ref{fig:BennuSims}. As mentioned, this model is sampled from the set of density distributions proposed by \citep{scheeres_heterogeneous_2020}. The density contrast of this anomaly and the background density $\rho_0$ are also constrained to be close to the density contrast of the ring and the background density of that same reference model. We select as weights of the level-set constraints (see Eq. \ref{eq:levelSetConstraints}) $\nu=\num{3}$, while for the density constraints the weight is 20 times larger. Figure \ref{fig:BennuReal_wConstraints} shows the corresponding GILA density solution, starting as usual from the generic initial distribution of Figure \ref{fig:referenceModels}. The $\chi_P^2$ is still close to 1, indicating a good fit of the data, as shown also by the RMS of the residuals for each gravity degree, with only degree-2 errors being larger than the measurements uncertainties (by a factor of about 2.5). The estimated model does present a lighter equatorial ring, as well as an inner region of low density, which was not enforced explicitly (although somewhat implicitly by constraining the background density to be larger than the bulk density). The density contrast of the ring is 35\% smaller in magnitude than that of the ring in the reference model (\qty{-400}{kg/m^3}), and the bulk density 13\% smaller than the value of the constraint (\qty{1370}{kg/m^3}). Therefore, the constraints added to the inversion do not appear excessively tight.

\section{Discussion}
\label{sec:Discussion}
Overall, simulations have proved that GILA is able to retrieve a target model in ideal conditions of noise-free data and perfect shape knowledge, notwithstanding the non-uniqueness of the gravity inversion. Typical target models where the method fails are those where the degeneracy is higher, such as the case of concentric shells (not shown here), which would require adapting the method. Similarly, we don't expect the method to correctly retrieve size and density jump of perfectly spherical anomalies, as their contribution to the gravity field is that of a point mass, although generally our tests still show a good retrieval of the total mass of the anomaly in such cases. \\

We tested the robustness of the inversion approach with respect to various possible deviations from such ideal settings, primarily a lower resolution of the observed gravitational potential and presence of noise on the measurements and the shape of the body. For the more irregular bodies the performance of GILA was consistently better than for more spherical bodies like Bennu and Phobos. For these latter two, the range of applicability of the method seems to be right at the edge of what might actually be expected in many real-mission scenarios, meaning a gravity field of degree 3 and uncertainty on the shape of about 0.5\% of the body radius. Nonetheless, we stress here that low values of the correlation metric used in this study may not mean a completely wrong model compared to the truth, as seen for example in Figure \ref{fig:vsGrav}. These limits can therefore be considered conservative, and in any case it is complicated to give quantitative expectations in terms of predicted performance in such an ill-posed problem. 

Still, a single density solution accounts for very little in the context of gravity inversion (unless strong constraints are added), since as of yet GILA has no way to provide realistic uncertainties associated with the estimated parameters. A possible approach, given the too-optimistic uncertainties output by the least squares, would be to compute the uncertainties on the converged solution via Markov-Chain Monte Carlo stochastic sampling about the estimated parameters, as in \citet{galleyGeophysicalInversion3D2020}. Even then, uncertainties on a single solution fitting the measurements would not be representative of the other families of interior equally agreeing with the data within the noise. We therefore deem it necessary to use the inversion algorithm of GILA multiple times with different inputs, to then extract statistics from the sets of solutions thus generated. The proposed approach, including a clustering of the solutions, and still computationally feasible for a number of solutions in the order of thousands, would in theory not only provide information on the uncertainties associated with a single family of solutions, but also an exploration of the range of possible interior distributions. While this exploration of the solution space is bound to be limited by computational constraints, we find that in the cases presented here it is excessively narrow. In most of our applications of the clustering approach, we detected less than 3 significantly different families of interior, although generally the true solution was included in one of those. This was confirmed by visual inspection, meaning that using higher-order moments of the density function would have meant separating solutions very similar to each other. This is certainly a shortcoming for a method ideated for an early exploration phase of the interior, when additional constraints are not yet available. Hence, additional efforts are required in removing biases from the random inputs and from the algorithm itself. \\

Another important limitation of the method in its current state is the residuals being limited by the model noise more often than the data noise. In applications over very precise measurements, this would mean reducing the information provided by the data and increasing the degeneracy of the problem. As the model error was a limiting factor in many of the tests shown here, the conclusions here derived could be seen as conservative. That is, assuming that improvements on GILA aimed at reducing these convergence issues (such as more accurate partials, better optimization algorithm, or forward optimization starting from the converged model) are effectively possible.

It is challenging to quantitatively compare the performance of GILA to any of the other approaches to the gravity inversion of small bodies mentioned in Section \ref{sec:introduction}, nor has this kind of comparison with different approaches over the same test cases been performed as of yet. As mentioned, the piecewise-constant assumption certainly makes this method more suitable than others for non-layered bodies which are indeed composed of distinct anomalies, be it because of fracturing or reaccretion. Nevertheless, as shown in Section \ref{sec:PhobosPC}, GILA can still retrieve solution that approximate simple smooth distributions. With respect to the forward approach to the shape determination of \citet{takahashi_morphology_2014}, the direct estimation of the anomaly shape as part of the least squares inversion leads perhaps to a more immediate solution, but at the cost of dealing with the pitfalls of non-convex estimation. 
On the other hand, methods such as that of \citet{tricarico_global_2013} are better suited than GILA if the density distribution is smooth, as well as providing a more complete view of the space of possible solutions, while the power of neural networks as universal function approximators makes the method of \citet{izzo_geodesy_2022} able to retrieve virtually any kind of interior distribution. In any case, when dealing with bodies about which little to nothing is known, synergy among different estimation approaches is essential, as demonstrated in \citet{scheeres_heterogeneous_2020}. \\

As for the application of GILA to OSIRIS-REx data, the flat residuals show good convergence properties of the method in real applications. The disagreement between the distribution output by GILA and the more plausible one proposed by \citet{scheeres_heterogeneous_2020} is realistically explained by the lack of physical constraints in our inversion, along with our apparent incomplete exploration of the solution space and the inherent non-uniqueness of the problem. This is confirmed by the convergence to something closer to the ring-core model as soon as direct constraints are added.

%The ability of our approach to accurately reconstruct the shapes of the anomalies quickly degrades as soon as realistic noise is added, although the correlations usually remain above 0.3 for noise up to 100\%, meaning that the inverted model is correctly representing the position and sign of the density jump of the target anomalies. \slm{(WE MAY ADD QUATITATIVE INFORMAITON ABOUT THE DEGRADATION, REFER ALSO TO APPENDIX, and generally speaking try to add positive sentences about the method)}\\
%
%The real-data solution shown in Figure \ref{fig:BennuReal}, while consistent with the gravity data and their formal errors, is not in agreement with the model proposed by \citet{scheeres_heterogeneous_2020}, consisting of underdense equatorial ring and core region. The disagreement is however explained by the lack of physical constraints in our inversion, and the inherent non-uniqueness of the problem.

\section{Conclusions}
We have presented GILA, a novel gravity inversion algorithm extending methods established in Earth geodesy to the gravity inversion of small irregular celestial bodies. The interior of the body is assumed to be divisible into a small number of regions where the density is uniform. The shape of each of these mass anomalies is represented implicitly by a level-set function, which is adjusted along with the density contrasts and the background density to fit the measured gravitational potential via iterative least-squares.
We have tested the algorithm over different synthetic scenarios, increasingly more realistic, and found it to provide reasonable approximations of the true model in most of the cases. The non-uniqueness of the problem is addressed by performing several inversions with different initial conditions. Yet one of the main limitations of the method in its current state is perhaps the limited exploration of the solution space it provides, which is essential for asteroids or comets, where most of the time little is known about the interior. As shown in the processing of Bennu real data, this can be mitigated by providing additional information to the method. This would imply extending GILA to the processing of different kinds of measurements, such as local gravimetric data or observations about the dynamical state of the body. If not by joint inversion, inputs from other spacecraft instruments or from theory should be added as constraints. For now, the only type of constraints tested are those directly on the target shape and density, but extension to general, non-quadratic constraints of different nature would be needed to increase the reliability of GILA's outputs in real scenarios.

\section{Acknowledgements}
This work relies heavily on open-source software, particularly the Python libraries \emph{NumPy} and \emph{PyVista}, the latter employed for mesh manipulation and some of the plots. Most of the figures presented here were generated using the \emph{Plotly} library, along with \emph{matplotlib} and \emph{seaborn}. We thank the developers for making their work easily accessible to the community. We also thank Daniel Scheeres, Matthias Noeker, and Grégoire Henry for their suggestions and feedback, as well as Véronique Dehant, Özgür Karatekin, Tim Van Hoolst, and François Massonnet, who supervise the project. This work is funded by the French community of Belgium, within the frame of a FRIA grant.

\section{Data Availability}
The code for GILA, which was used to generate the data presented in this paper, is available on GitLab \footnote{\url{https://gitlab-as.oma.be/sbm/gila}}. 

% \bibliography{levelSet}
\bibliographystyle{tudelft-report}
\bibliography{levelSet}

\begin{thebibliography}{63}
\providecommand{\natexlab}[1]{#1}
\providecommand{\url}[1]{\texttt{#1}}
\expandafter\ifx\csname urlstyle\endcsname\relax
  \providecommand{\doi}[1]{doi: #1}\else
  \providecommand{\doi}{doi: \begingroup \urlstyle{rm}\Url}\fi

\bibitem[Adalsteinsson and Sethian(1995)]{adalsteinssonFastLevelSet1995}
D.~Adalsteinsson and J.~A. Sethian.
\newblock A {{Fast Level Set Method}} for {{Propagating Interfaces}}.
\newblock \emph{Journal of Computational Physics}, 118\penalty0 (2):\penalty0
  269--277, May 1995.
\newblock ISSN 0021-9991.
\newblock \doi{10.1006/jcph.1995.1098}.

\bibitem[Asad et~al.(2021)Asad, Philpott, Johnson, Barnouin, Palmer, Weirich,
  Daly, Perry, Gaskell, Bierhaus, Seabrook, Espiritu, Nair, Ernst, Daly, Nolan,
  Enos, and Lauretta]{asadValidationStereophotoclinometricShape2021}
M.~M.~A. Asad, L.~C. Philpott, C.~L. Johnson, et~al.
\newblock Validation of {{Stereophotoclinometric Shape Models}} of {{Asteroid}}
  (101955) {{Bennu}} during the {{OSIRIS-REx Mission}}.
\newblock \emph{The Planetary Science Journal}, 2\penalty0 (2):\penalty0 82,
  April 2021.
\newblock ISSN 2632-3338.
\newblock \doi{10.3847/PSJ/abe4dc}.

\bibitem[Bj{\"o}rck(2015)]{bjorckNumericalMethodsMatrix2015}
{\AA}.~Bj{\"o}rck.
\newblock \emph{Numerical {{Methods}} in {{Matrix Computations}}}, volume~59 of
  \emph{Texts in {{Applied Mathematics}}}.
\newblock {Springer International Publishing}, {Cham}, 2015.
\newblock ISBN 978-3-319-05088-1 978-3-319-05089-8.
\newblock \doi{10.1007/978-3-319-05089-8}.

\bibitem[Blakely(1995)]{blakelyPotentialTheoryGravity1995}
R.~J. Blakely.
\newblock \emph{Potential {{Theory}} in {{Gravity}} and {{Magnetic
  Applications}}}.
\newblock {Cambridge University Press}, {Cambridge}, 1995.
\newblock ISBN 978-0-521-41508-8.
\newblock \doi{10.1017/CBO9780511549816}.

\bibitem[Carry(2012)]{carryDensityAsteroids2012}
B.~Carry.
\newblock Density of asteroids.
\newblock \emph{Planetary and Space Science}, 73\penalty0 (1):\penalty0
  98--118, December 2012.
\newblock ISSN 0032-0633.
\newblock \doi{10.1016/j.pss.2012.03.009}.

\bibitem[Chao(2005)]{chaoInversionMassDistribution2005}
B.~F. Chao.
\newblock On inversion for mass distribution from global (time-variable)
  gravity field.
\newblock \emph{Journal of Geodynamics}, 39\penalty0 (3):\penalty0 223--230,
  April 2005.
\newblock ISSN 0264-3707.
\newblock \doi{10.1016/j.jog.2004.11.001}.

\bibitem[Chesley et~al.(2020)Chesley, French, Davis, Jacobson, Brozovi{\'c},
  Farnocchia, Selznick, Liounis, Hergenrother, Moreau, Pelgrift,
  {Lessac-Chenen}, Molaro, Park, Rozitis, Scheeres, Takahashi,
  Vokrouhlick{\'y}, Wolner, Adam, Bos, Christensen, Emery, Leonard, McMahon,
  Nolan, Shelly, and Lauretta]{chesleyTrajectoryEstimationParticles2020}
S.~R. Chesley, A.~S. French, A.~B. Davis, et~al.
\newblock Trajectory {{Estimation}} for {{Particles Observed}} in the
  {{Vicinity}} of (101955) {{Bennu}}.
\newblock \emph{Journal of Geophysical Research: Planets}, 125\penalty0
  (9):\penalty0 e2019JE006363, 2020.
\newblock ISSN 2169-9100.
\newblock \doi{10.1029/2019JE006363}.

\bibitem[Dinsmore and
  {de~Wit}(2023)]{dinsmoreConstrainingInteriorsAsteroids2023}
J.~T. Dinsmore and J.~{de~Wit}.
\newblock Constraining the interiors of asteroids through close encounters.
\newblock \emph{Monthly Notices of the Royal Astronomical Society},
  520\penalty0 (3):\penalty0 3459--3475, April 2023.
\newblock ISSN 0035-8711.
\newblock \doi{10.1093/mnras/stac2866}.

\bibitem[Ermakov et~al.(2017)Ermakov, Fu, {Castillo-Rogez}, Raymond, Park,
  Preusker, Russell, Smith, and Zuber]{ermakovConstraintsCeresInternal2017}
A.~I. Ermakov, R.~R. Fu, J.~C. {Castillo-Rogez}, et~al.
\newblock Constraints on {{Ceres}}' {{Internal Structure}} and {{Evolution From
  Its Shape}} and {{Gravity Measured}} by the {{Dawn Spacecraft}}.
\newblock \emph{Journal of Geophysical Research: Planets}, 122\penalty0
  (11):\penalty0 2267--2293, 2017.
\newblock ISSN 2169-9100.
\newblock \doi{10.1002/2017JE005302}.

\bibitem[Ermakov et~al.(2014)Ermakov, Zuber, Smith, Raymond, Balmino, Fu, and
  Ivanov]{ermakovConstraintsVestaInterior2014}
A.~I. Ermakov, M.~T. Zuber, D.~E. Smith, et~al.
\newblock Constraints on {{Vesta}}'s interior structure using gravity and shape
  models from the {{Dawn}} mission.
\newblock \emph{Icarus}, 240:\penalty0 146--160, September 2014.
\newblock ISSN 0019-1035.
\newblock \doi{10.1016/j.icarus.2014.05.015}.

\bibitem[Fujiwara et~al.(2006)Fujiwara, Kawaguchi, Yeomans, Abe, Mukai, Okada,
  Saito, Yano, Yoshikawa, Scheeres, {Barnouin-Jha}, Cheng, Demura, Gaskell,
  Hirata, Ikeda, Kominato, Miyamoto, Nakamura, Nakamura, Sasaki, and
  Uesugi]{fujiwaraRubblePileAsteroidItokawa2006}
A.~Fujiwara, J.~Kawaguchi, D.~K. Yeomans, et~al.
\newblock The {{Rubble-Pile Asteroid Itokawa}} as {{Observed}} by {{Hayabusa}}.
\newblock \emph{Science}, 312\penalty0 (5778):\penalty0 1330--1334, June 2006.
\newblock \doi{10.1126/science.1125841}.

\bibitem[Galley et~al.(2020)Galley, Leli{\`e}vre, and
  Farquharson]{galleyGeophysicalInversion3D2020}
C.~G. Galley, P.~G. Leli{\`e}vre, and C.~G. Farquharson.
\newblock Geophysical inversion for {{3D}} contact surface geometry.
\newblock \emph{GEOPHYSICS}, 85\penalty0 (6):\penalty0 K27--K45, November 2020.
\newblock ISSN 0016-8033.
\newblock \doi{10.1190/geo2019-0614.1}.

\bibitem[Giraud et~al.(2021)Giraud, Lindsay, and
  Jessell]{giraud_generalization_2021}
J.~Giraud, M.~Lindsay, and M.~Jessell.
\newblock Generalization of level-set inversion to an arbitrary number of
  geologic units in a regularized least-squares framework.
\newblock \emph{GEOPHYSICS}, 86\penalty0 (4):\penalty0 R623--R637, July 2021.
\newblock ISSN 0016-8033.
\newblock \doi{10.1190/geo2020-0263.1}.

\bibitem[Goossens and Smith(2023)]{goossensGravityDegreeDepth2023}
S.~Goossens and D.~E. Smith.
\newblock Gravity degree\textendash depth relationship using point mass
  spherical harmonics.
\newblock \emph{Geophysical Journal International}, 233\penalty0 (3):\penalty0
  1878--1889, June 2023.
\newblock ISSN 0956-540X.
\newblock \doi{10.1093/gji/ggad036}.

\bibitem[Hedges et~al.(2017)Hedges, Kim, and
  Jack]{hedgesStochasticLevelsetMethod2017}
L.~O. Hedges, H.~A. Kim, and R.~L. Jack.
\newblock Stochastic level-set method for shape optimisation.
\newblock \emph{Journal of Computational Physics}, 348:\penalty0 82--107,
  November 2017.
\newblock ISSN 00219991.
\newblock \doi{10.1016/j.jcp.2017.07.010}.

\bibitem[Heiskanen and Moritz(1967)]{heiskanen1967physical}
W.~Heiskanen and H.~Moritz.
\newblock \emph{Physical Geodesy}.
\newblock Series of Books in Geology. {W. H. Freeman}, 1967.
\newblock ISBN 978-0-608-30923-1.

\bibitem[Herique et~al.(2022)Herique, Plettemeier, and
  Kofman]{heriqueJuRaJuventasRadar2022}
A.~Herique, D.~Plettemeier, and W.~W. Kofman.
\newblock {{JuRa}}: The {{Juventas Radar}} on {{Hera}} to fathom {{Dimorphos}}.
\newblock 2022:\penalty0 NH12C--0296, December 2022.

\bibitem[Izquierdo et~al.(2023)Izquierdo, Leki{\'c}, and
  Mont{\'e}si]{izquierdoObjectOrientedBayesianGravity2023}
K.~Izquierdo, V.~Leki{\'c}, and L.~G.~J. Mont{\'e}si.
\newblock An {{Object-Oriented Bayesian Gravity Inversion Scheme}} for
  {{Inferring Density Anomalies}} in {{Planetary Interiors}}.
\newblock \emph{Earth and Space Science}, 10\penalty0 (7):\penalty0
  e2023EA002853, 2023.
\newblock ISSN 2333-5084.
\newblock \doi{10.1029/2023EA002853}.

\bibitem[Izzo and G{\'o}mez(2022)]{izzo_geodesy_2022}
D.~Izzo and P.~G{\'o}mez.
\newblock Geodesy of irregular small bodies via neural density fields.
\newblock \emph{Communications Engineering}, 1\penalty0 (1):\penalty0 1--12,
  December 2022.
\newblock ISSN 2731-3395.
\newblock \doi{10.1038/s44172-022-00050-3}.

\bibitem[Jamet and Tsoulis(2020)]{jametLineIntegralApproach2020}
O.~Jamet and D.~Tsoulis.
\newblock A line integral approach for the computation of the potential
  harmonic coefficients of a constant density polyhedron.
\newblock \emph{Journal of Geodesy}, 94\penalty0 (3):\penalty0 30, 2020.

\bibitem[Jekeli(2007)]{jekeliPotentialTheoryStatic2007}
C.~Jekeli.
\newblock \emph{Potential {{Theory}} and {{Static Gravity Field}} of the
  {{Earth}}}, volume~3.
\newblock January 2007.
\newblock \doi{10.1016/B978-044452748-6.00054-7}.

\bibitem[Kofman et~al.(2020)Kofman, Zine, Herique, Rogez, Jorda, and
  {Levasseur-Regourd}]{kofmanInteriorComet67P2020}
W.~Kofman, S.~Zine, A.~Herique, et~al.
\newblock The interior of {{Comet 67P}}/{{C}}\textendash{{G}}; revisiting
  {{CONSERT}} results with the exact position of the {{Philae}} lander.
\newblock \emph{Monthly Notices of the Royal Astronomical Society},
  497\penalty0 (3):\penalty0 2616--2622, September 2020.
\newblock ISSN 0035-8711.
\newblock \doi{10.1093/mnras/staa2001}.

\bibitem[Konopliv et~al.(2014)Konopliv, Asmar, Park, Bills, Centinello,
  Chamberlin, Ermakov, Gaskell, Rambaux, Raymond, Russell, Smith, Tricarico,
  and Zuber]{konopliv_vesta_2014}
A.~S. Konopliv, S.~W. Asmar, R.~S. Park, et~al.
\newblock The {{Vesta}} gravity field, spin pole and rotation period, landmark
  positions, and ephemeris from the {{Dawn}} tracking and optical data.
\newblock \emph{Icarus}, 240:\penalty0 103--117, September 2014.
\newblock ISSN 0019-1035.
\newblock \doi{10.1016/j.icarus.2013.09.005}.

\bibitem[Konopliv et~al.(2018)Konopliv, Park, Vaughan, Bills, Asmar, Ermakov,
  Rambaux, Raymond, {Castillo-Rogez}, Russell, Smith, and
  Zuber]{konopliv_ceres_2018}
A.~S. Konopliv, R.~S. Park, A.~T. Vaughan, et~al.
\newblock The {{Ceres}} gravity field, spin pole, rotation period and orbit
  from the {{Dawn}} radiometric tracking and optical data.
\newblock \emph{Icarus}, 299:\penalty0 411--429, January 2018.
\newblock ISSN 0019-1035.
\newblock \doi{10.1016/j.icarus.2017.08.005}.

\bibitem[Le~Maistre et~al.(2019)Le~Maistre, Rivoldini, and
  Rosenblatt]{le_maistre_signature_2019}
S.~Le~Maistre, A.~Rivoldini, and P.~Rosenblatt.
\newblock Signature of {Phobos}’ interior structure in its gravity field and
  libration.
\newblock \emph{Icarus}, 321:\penalty0 272--290, 2019.
\newblock ISSN 0019-1035.
\newblock \doi{10.1016/j.icarus.2018.11.022}.

\bibitem[Ledbetter et~al.(2021)Ledbetter, Sood, Keane, and
  Stuart]{ledbetterSmallSatSwarmGravimetry2021}
W.~G. Ledbetter, R.~Sood, J.~Keane, and J.~Stuart.
\newblock {{SmallSat}} swarm gravimetry: {{Revealing}} the interior structure
  of asteroids and comets.
\newblock \emph{Astrodynamics}, 5\penalty0 (3):\penalty0 217--236, September
  2021.
\newblock ISSN 2522-0098.
\newblock \doi{10.1007/s42064-020-0098-1}.

\bibitem[Lee~Rodgers and Nicewander(1988)]{leerodgersThirteenWaysLook1988}
J.~Lee~Rodgers and W.~A. Nicewander.
\newblock Thirteen {{Ways}} to {{Look}} at the {{Correlation Coefficient}}.
\newblock \emph{The American Statistician}, 42\penalty0 (1):\penalty0 59--66,
  February 1988.
\newblock ISSN 0003-1305.
\newblock \doi{10.1080/00031305.1988.10475524}.

\bibitem[Levison et~al.(2021)Levison, Olkin, Noll, Marchi, Iii, Bierhaus,
  Binzel, Bottke, Britt, Brown, Buie, Christensen, Emery, Grundy, Hamilton,
  Howett, Mottola, P{\"a}tzold, Reuter, Spencer, Statler, Stern, Sunshine,
  Weaver, and Wong]{levisonLucyMissionTrojan2021}
H.~F. Levison, C.~B. Olkin, K.~S. Noll, et~al.
\newblock Lucy {{Mission}} to the {{Trojan Asteroids}}: {{Science Goals}}.
\newblock \emph{The Planetary Science Journal}, 2\penalty0 (5):\penalty0 171,
  August 2021.
\newblock ISSN 2632-3338.
\newblock \doi{10.3847/PSJ/abf840}.

\bibitem[Li et~al.(2017)Li, Lu, Qian, and Li]{li_multiple_2017}
W.~Li, W.~Lu, J.~Qian, and Y.~Li.
\newblock A multiple level set method for three-dimensional inversion of
  magnetic data.
\newblock \emph{GEOPHYSICS}, 82:\penalty0 1--92, June 2017.
\newblock \doi{10.1190/geo2016-0530.1}.

\bibitem[Li and Oldenburg(1998)]{li_3-d_1998}
Y.~Li and D.~W. Oldenburg.
\newblock 3-{{D}} inversion of gravity data.
\newblock \emph{GEOPHYSICS}, 63\penalty0 (1):\penalty0 109--119, January 1998.
\newblock ISSN 0016-8033, 1942-2156.
\newblock \doi{10.1190/1.1444302}.

\bibitem[Lowry et~al.(2014)Lowry, Weissman, Duddy, Rozitis, Fitzsimmons, Green,
  Hicks, Snodgrass, Wolters, Chesley, Pittichov{\'a}, and van
  Oers]{lowryInternalStructureAsteroid2014}
S.~C. Lowry, P.~R. Weissman, S.~R. Duddy, et~al.
\newblock The internal structure of asteroid (25143) {{Itokawa}} as revealed by
  detection of {{YORP}} spin-up.
\newblock \emph{Astronomy \& Astrophysics}, 562:\penalty0 A48, February 2014.
\newblock ISSN 0004-6361, 1432-0746.
\newblock \doi{10.1051/0004-6361/201322602}.

\bibitem[Matsumoto et~al.(2021)Matsumoto, Hirata, Ikeda, Kouyama, Senshu,
  Yamamoto, Noda, Miyamoto, Araya, Araki, Kamata, Baresi, and
  Namiki]{matsumotoMMXGeodesyInvestigations2021}
K.~Matsumoto, N.~Hirata, H.~Ikeda, et~al.
\newblock {{MMX}} geodesy investigations: Science requirements and observation
  strategy.
\newblock \emph{Earth, Planets and Space}, 73\penalty0 (1):\penalty0 226,
  December 2021.
\newblock ISSN 1880-5981.
\newblock \doi{10.1186/s40623-021-01500-6}.

\bibitem[Michel et~al.(2022)Michel, K{\"u}ppers, Bagatin, Carry, Charnoz,
  de~Leon, Fitzsimmons, Gordo, Green, H{\'e}rique, Juzi, Karatekin, Kohout,
  Lazzarin, Murdoch, Okada, Palomba, Pravec, Snodgrass, Tortora, Tsiganis,
  Ulamec, Vincent, W{\"u}nnemann, Zhang, Raducan, Dotto, Chabot, Cheng, Rivkin,
  Barnouin, Ernst, Stickle, Richardson, Thomas, Arakawa, Miyamoto, Nakamura,
  Sugita, Yoshikawa, Abell, Asphaug, Ballouz, Bottke, Lauretta, Walsh, Martino,
  and Carnelli]{michelESAHeraMission2022}
P.~Michel, M.~K{\"u}ppers, A.~C. Bagatin, et~al.
\newblock The {{ESA Hera Mission}}: {{Detailed Characterization}} of the {{DART
  Impact Outcome}} and of the {{Binary Asteroid}} (65803) {{Didymos}}.
\newblock \emph{The Planetary Science Journal}, 3\penalty0 (7):\penalty0 160,
  July 2022.
\newblock ISSN 2632-3338.
\newblock \doi{10.3847/PSJ/ac6f52}.

\bibitem[Michel and Fokas(2008)]{michel_unified_2008}
V.~Michel and A.~S. Fokas.
\newblock A unified approach to various techniques for the non-uniqueness of
  the inverse gravimetric problem and wavelet-based methods.
\newblock \emph{Inverse Problems}, 24\penalty0 (4):\penalty0 045019, July 2008.
\newblock ISSN 0266-5611.
\newblock \doi{10.1088/0266-5611/24/4/045019}.

\bibitem[Miller et~al.(2002)Miller, Konopliv, Antreasian, Bordi, Chesley,
  Helfrich, Owen, Wang, Williams, Yeomans, and
  Scheeres]{miller_determination_2002-1}
J.~Miller, A.~Konopliv, P.~Antreasian, et~al.
\newblock Determination of {{Shape}}, {{Gravity}}, and {{Rotational State}} of
  {{Asteroid}} 433 {{Eros}}.
\newblock \emph{Icarus}, 155\penalty0 (1):\penalty0 3--17, January 2002.
\newblock ISSN 00191035.
\newblock \doi{10.1006/icar.2001.6753}.

\bibitem[Osher and Fedkiw(2003)]{osherLevelSetMethods2003}
S.~Osher and R.~Fedkiw.
\newblock \emph{Level {{Set Methods}} and {{Dynamic Implicit Surfaces}}},
  volume 153 of \emph{Applied {{Mathematical Sciences}}}.
\newblock {Springer}, {New York, NY}, 2003.
\newblock ISBN 978-1-4684-9251-4 978-0-387-22746-7.
\newblock \doi{10.1007/b98879}.

\bibitem[Park et~al.(2014)Park, Konopliv, Asmar, Bills, Gaskell, Raymond,
  Smith, Toplis, and Zuber]{parkGravityFieldExpansion2014}
R.~S. Park, A.~S. Konopliv, S.~W. Asmar, et~al.
\newblock Gravity field expansion in ellipsoidal harmonic and polyhedral
  internal representations applied to {{Vesta}}.
\newblock \emph{Icarus}, 240:\penalty0 118--132, September 2014.
\newblock ISSN 0019-1035.
\newblock \doi{10.1016/j.icarus.2013.12.005}.

\bibitem[Park et~al.(2010)Park, Werner, and
  Bhaskaran]{parkEstimatingSmallBodyGravity2010}
R.~S. Park, R.~A. Werner, and S.~Bhaskaran.
\newblock Estimating {{Small-Body Gravity Field}} from {{Shape Model}} and
  {{Navigation Data}}.
\newblock \emph{Journal of Guidance, Control, and Dynamics}, 33\penalty0
  (1):\penalty0 212--221, January 2010.
\newblock ISSN 0731-5090.
\newblock \doi{10.2514/1.41585}.

\bibitem[P{\"a}tzold et~al.(2016)P{\"a}tzold, Andert, Hahn, Asmar, Barriot,
  Bird, H{\"a}usler, Peter, Tellmann, Gr{\"u}n, Weissman, Sierks, Jorda,
  Gaskell, Preusker, and Scholten]{patzoldHomogeneousNucleusComet2016}
M.~P{\"a}tzold, T.~Andert, M.~Hahn, et~al.
\newblock A homogeneous nucleus for comet
  {{67P}}/{{Churyumov}}\textendash{{Gerasimenko}} from its gravity field.
\newblock \emph{Nature}, 530\penalty0 (7588):\penalty0 63--65, February 2016.
\newblock ISSN 1476-4687.
\newblock \doi{10.1038/nature16535}.

\bibitem[Pedregosa et~al.(2011)Pedregosa, Varoquaux, Gramfort, Michel, Thirion,
  Grisel, Blondel, Prettenhofer, Weiss, Dubourg, Vanderplas, Passos,
  Cournapeau, Brucher, Perrot, and Duchesnay]{scikit-learn}
F.~Pedregosa, G.~Varoquaux, A.~Gramfort, et~al.
\newblock Scikit-learn: Machine learning in {P}ython.
\newblock \emph{Journal of Machine Learning Research}, 12:\penalty0 2825--2830,
  2011.

\bibitem[{Perez-Molina} et~al.(2022){Perez-Molina}, {Campo-Bagatin}, and
  Tr{\'o}golo]{perez-molinaFFTGravityField2022a}
M.~{Perez-Molina}, A.~{Campo-Bagatin}, and N.~Tr{\'o}golo.
\newblock {{FFT}} gravity field calculation method and super-ellipsoid
  generated field.
\newblock Technical Report EPSC2022-995, {Copernicus Meetings}, July 2022.

\bibitem[Ritter et~al.(2022)Ritter, Karatekin, Carrasco, Tasev,
  Alav{\'e}s~Ma{\~n}ogil, Noeker, Van~Ransbeek, Noiset, and
  Berk~Senel]{ritterMeasuringGravityGRASS2022a}
B.~Ritter, {\"O}.~Karatekin, J.~A. Carrasco, et~al.
\newblock Measuring gravity with the {{GRASS}} instrument on the {{Hera}}
  mission.
\newblock pages EPSC2022--1115, September 2022.
\newblock \doi{10.5194/epsc2022-1115}.

\bibitem[R\"ucker et~al.(2017)R\"ucker, G\"unther, and Wagner]{Ruecker2017}
C.~R\"ucker, T.~G\"unther, and F.~M. Wagner.
\newblock {pyGIMLi}: An open-source library for modelling and inversion in
  geophysics.
\newblock \emph{Computers and Geosciences}, 109:\penalty0 106--123, 2017.
\newblock \doi{10.1016/j.cageo.2017.07.011}.

\bibitem[Scheeres et~al.(2000)Scheeres, Khushalani, and
  Werner]{scheeres_estimating_2000}
D.~J. Scheeres, B.~Khushalani, and R.~A. Werner.
\newblock Estimating asteroid density distributions from shape and gravity
  information.
\newblock \emph{Planetary and Space Science}, 48\penalty0 (10):\penalty0
  965--971, 2000.
\newblock ISSN 0032-0633.
\newblock \doi{10.1016/S0032-0633(00)00064-7}.

\bibitem[Scheeres et~al.(2015)Scheeres, Britt, Carry, and
  Holsapple]{scheeresAsteroidInteriorsMorphology2015a}
D.~J. Scheeres, D.~Britt, B.~Carry, and K.~A. Holsapple.
\newblock Asteroid {{Interiors}} and {{Morphology}}.
\newblock In P.~Michel, F.~E. DeMeo, and W.~F. Bottke, editors, \emph{Asteroids
  {{IV}}}. {University of Arizona Press}, 2015.
\newblock ISBN 978-0-8165-3213-1.
\newblock \doi{10.2458/azu_uapress_9780816532131-ch038}.

\bibitem[Scheeres et~al.(2016)Scheeres, Hesar, Tardivel, Hirabayashi,
  Farnocchia, McMahon, Chesley, Barnouin, Binzel, Bottke, Daly, Emery,
  Hergenrother, Lauretta, Marshall, Michel, Nolan, and
  Walsh]{scheeres_geophysical_2016}
D.~J. Scheeres, S.~Hesar, S.~Tardivel, et~al.
\newblock The geophysical environment of {{Bennu}}.
\newblock \emph{Icarus}, 276:\penalty0 116--140, September 2016.
\newblock ISSN 00191035.
\newblock \doi{10.1016/j.icarus.2016.04.013}.

\bibitem[Scheeres et~al.(2020)Scheeres, French, Tricarico, Chesley, Takahashi,
  Farnocchia, McMahon, Brack, Davis, Ballouz, Jawin, Rozitis, Emery, Ryan,
  Park, Rush, Mastrodemos, Kennedy, Bellerose, Lubey, Velez, Vaughan, Leonard,
  Geeraert, Page, Antreasian, Mazarico, Getzandanner, Rowlands, Moreau, Small,
  Highsmith, Goossens, Palmer, Weirich, Gaskell, Barnouin, Daly, Seabrook,
  Asad, Philpott, Johnson, Hartzell, Hamilton, Michel, Walsh, Nolan, and
  Lauretta]{scheeres_heterogeneous_2020}
D.~J. Scheeres, A.~S. French, P.~Tricarico, et~al.
\newblock Heterogeneous mass distribution of the rubble-pile asteroid (101955)
  {{Bennu}}.
\newblock \emph{Science Advances}, 6\penalty0 (41):\penalty0 eabc3350, October
  2020.
\newblock ISSN 2375-2548.
\newblock \doi{10.1126/sciadv.abc3350}.

\bibitem[Sethian(1996)]{sethianFastMarchingLevel1996}
J.~A. Sethian.
\newblock A fast marching level set method for monotonically advancing fronts.
\newblock \emph{Proceedings of the National Academy of Sciences of the United
  States of America}, 93\penalty0 (4):\penalty0 1591--1595, February 1996.
\newblock ISSN 0027-8424.

\bibitem[Silva and Barbosa(2006)]{silvaInteractiveGravityInversion2006}
J.~B.~C. Silva and V.~C.~F. Barbosa.
\newblock Interactive gravity inversion.
\newblock \emph{GEOPHYSICS}, 71\penalty0 (1):\penalty0 J1--J9, January 2006.
\newblock ISSN 0016-8033, 1942-2156.
\newblock \doi{10.1190/1.2168010}.

\bibitem[Sorsa et~al.(2020)Sorsa, Takala, Bambach, Deller, Vilenius, Agarwal,
  Carroll, Karatekin, and Pursiainen]{sorsa_tomographic_2020}
L.-I. Sorsa, M.~Takala, P.~Bambach, et~al.
\newblock Tomographic inversion of gravity gradient field for a synthetic
  {Itokawa} model.
\newblock \emph{Icarus}, 336:\penalty0 113425, 2020.
\newblock ISSN 00191035.
\newblock \doi{10.1016/j.icarus.2019.113425}.

\bibitem[Takahashi and Scheeres(2014{\natexlab{a}})]{takahashi_morphology_2014}
Y.~Takahashi and D.~J. Scheeres.
\newblock Morphology driven density distribution estimation for small bodies.
\newblock \emph{Icarus}, 233:\penalty0 179--193, 2014{\natexlab{a}}.
\newblock ISSN 0019-1035.
\newblock \doi{10.1016/j.icarus.2014.02.004}.

\bibitem[Takahashi and Scheeres(2014{\natexlab{b}})]{takahashi_small_2014}
Y.~Takahashi and D.~J. Scheeres.
\newblock Small body surface gravity fields via spherical harmonic expansions.
\newblock \emph{Celestial Mechanics and Dynamical Astronomy}, 119\penalty0
  (2):\penalty0 169--206, June 2014{\natexlab{b}}.
\newblock ISSN 1572-9478.
\newblock \doi{10.1007/s10569-014-9552-9}.

\bibitem[Tardivel(2016)]{tardivel_limits_2016}
S.~Tardivel.
\newblock The {{Limits}} of the {{Mascons Approximation}} of the {{Homogeneous
  Polyhedron}}.
\newblock In \emph{{{AIAA}}/{{AAS Astrodynamics Specialist Conference}}},
  {{AIAA SPACE Forum}}. {American Institute of Aeronautics and Astronautics},
  September 2016.
\newblock \doi{10.2514/6.2016-5261}.

\bibitem[Tricarico et~al.(2021)Tricarico, Scheeres, French, McMahon, Brack,
  Leonard, Antreasian, Chesley, Farnocchia, Takahashi, Mazarico, Rowlands,
  Highsmith, Getzandanner, Moreau, Johnson, Philpott, Bierhaus, Walsh,
  Barnouin, Palmer, Weirich, Gaskell, Daly, Seabrook, Nolan, and
  Lauretta]{tricaricoInternalRubbleProperties2021}
P.~Tricarico, D.~J. Scheeres, A.~S. French, et~al.
\newblock Internal rubble properties of asteroid (101955) {{Bennu}}.
\newblock \emph{Icarus}, 370:\penalty0 114665, December 2021.
\newblock ISSN 0019-1035.
\newblock \doi{10.1016/j.icarus.2021.114665}.

\bibitem[Tricarico(2013)]{tricarico_global_2013}
P.~Tricarico.
\newblock Global gravity inversion of bodies with arbitrary shape.
\newblock \emph{Geophysical Journal International}, 195\penalty0 (1):\penalty0
  260--275, 2013.
\newblock ISSN 0956-540X.
\newblock \doi{10.1093/gji/ggt268}.

\bibitem[Tsoulis et~al.(2009)Tsoulis, Jamet, Verdun, and
  Gonindard]{tsoulis_recursive_2009}
D.~Tsoulis, O.~Jamet, J.~Verdun, and N.~Gonindard.
\newblock Recursive algorithms for the computation of the potential harmonic
  coefficients of a constant density polyhedron.
\newblock \emph{Journal of Geodesy}, 83\penalty0 (10):\penalty0 925--942,
  October 2009.
\newblock ISSN 0949-7714, 1432-1394.
\newblock \doi{10.1007/s00190-009-0310-9}.

\bibitem[Walsh(2018)]{walshRubblePileAsteroids2018}
K.~J. Walsh.
\newblock Rubble {{Pile Asteroids}}.
\newblock \emph{Annual Review of Astronomy and Astrophysics}, 56\penalty0
  (1):\penalty0 593--624, 2018.
\newblock \doi{10.1146/annurev-astro-081817-052013}.

\bibitem[Watanabe et~al.(2019)Watanabe, Hirabayashi, Hirata, Hirata, Noguchi,
  Shimaki, Ikeda, Tatsumi, Yoshikawa, Kikuchi, Yabuta, Nakamura, Tachibana,
  Ishihara, Morota, Kitazato, Sakatani, Matsumoto, Wada, Senshu, Honda,
  Michikami, Takeuchi, Kouyama, Honda, Kameda, Fuse, Miyamoto, Komatsu, Sugita,
  Okada, Namiki, Arakawa, Ishiguro, Abe, Gaskell, Palmer, Barnouin, Michel,
  French, McMahon, Scheeres, Abell, Yamamoto, Tanaka, Shirai, Matsuoka, Yamada,
  Yokota, Suzuki, Yoshioka, Cho, Tanaka, Nishikawa, Sugiyama, Kikuchi, Hemmi,
  Yamaguchi, Ogawa, Ono, Mimasu, Yoshikawa, Takahashi, Takei, Fujii, Hirose,
  Iwata, Hayakawa, Hosoda, Mori, Sawada, Shimada, Soldini, Yano, Tsukizaki,
  Ozaki, Iijima, Ogawa, Fujimoto, Ho, Moussi, Jaumann, Bibring, Krause, Terui,
  Saiki, Nakazawa, and Tsuda]{watanabeHayabusa2ArrivesCarbonaceous2019a}
S.~Watanabe, M.~Hirabayashi, N.~Hirata, et~al.
\newblock Hayabusa2 arrives at the carbonaceous asteroid 162173 {{Ryugu-A}}
  spinning top-shaped rubble pile.
\newblock \emph{Science (New York, N.Y.)}, 364\penalty0 (6437):\penalty0
  268--272, April 2019.
\newblock ISSN 1095-9203.
\newblock \doi{10.1126/science.aav8032}.

\bibitem[Werner(1997)]{werner_spherical_1997}
R.~A. Werner.
\newblock Spherical harmonic coefficients for the potential of a
  constant-density polyhedron.
\newblock \emph{Computers \& Geosciences}, 23\penalty0 (10):\penalty0
  1071--1077, December 1997.
\newblock ISSN 0098-3004.
\newblock \doi{10.1016/S0098-3004(97)00110-6}.

\bibitem[Willner et~al.(2014)Willner, Shi, and
  Oberst]{willnerPhobosShapeTopography2014a}
K.~Willner, X.~Shi, and J.~Oberst.
\newblock Phobos' shape and topography models.
\newblock \emph{Planetary and Space Science}, 102:\penalty0 51--59, November
  2014.
\newblock ISSN 0032-0633.
\newblock \doi{10.1016/j.pss.2013.12.006}.

\bibitem[Zhang et~al.(2022)Zhang, Michel, Barnouin, Roberts, Daly, Ballouz,
  Walsh, Richardson, Hartzell, and
  Lauretta]{zhangInferringInteriorsStructural2022}
Y.~Zhang, P.~Michel, O.~S. Barnouin, et~al.
\newblock Inferring interiors and structural history of top-shaped asteroids
  from external properties of asteroid (101955) {{Bennu}}.
\newblock \emph{Nature Communications}, 13\penalty0 (1):\penalty0 4589, August
  2022.
\newblock ISSN 2041-1723.
\newblock \doi{10.1038/s41467-022-32288-y}.

\bibitem[Zheglova et~al.(2017)Zheglova, Lelievre, and
  Farquharson]{zheglova_multiple_2017}
P.~Zheglova, P.~Lelievre, and C.~Farquharson.
\newblock Multiple level-set joint inversion of traveltime and gravity data
  with application to ore delineation: {{A}} synthetic study.
\newblock \emph{GEOPHYSICS}, 83:\penalty0 1--101, October 2017.
\newblock \doi{10.1190/geo2016-0675.1}.

\bibitem[Zuber et~al.(2022)Zuber, Park, {Elkins-Tanton}, Bell, Bruvold,
  Bercovici, Bills, Binzel, Jaumann, Marchi, Raymond, Roatsch, Wang, Weiss,
  Wenkert, and Wieczorek]{zuberPsycheGravityInvestigation2022}
M.~T. Zuber, R.~S. Park, L.~T. {Elkins-Tanton}, et~al.
\newblock The {{Psyche Gravity Investigation}}.
\newblock \emph{Space Science Reviews}, 218\penalty0 (8):\penalty0 57, October
  2022.
\newblock ISSN 1572-9672.
\newblock \doi{10.1007/s11214-022-00905-3}.

\end{thebibliography}
\end{document}